\documentclass[journal,onecolumn]{IEEEtran}

\usepackage{epsfig,color,amsmath,cite}
\usepackage{amsthm} 
\usepackage{amsmath}    
\usepackage[T1]{fontenc}
\usepackage[utf8]{inputenc}
\IEEEoverridecommandlockouts
\usepackage{bm}
\usepackage{epstopdf}
\usepackage{amssymb}
\usepackage{url}
\usepackage{enumitem} 
\usepackage{multirow}
\usepackage{hhline}
\usepackage{booktabs}
\usepackage{mathtools}
\usepackage{makecell}
\usepackage{empheq}
\usepackage[linesnumbered,boxed,commentsnumbered,ruled,vlined,longend]{algorithm2e}
\usepackage{comment}

\makeatother
\DeclareMathAlphabet\mathbfcal{OMS}{cmsy}{b}{n}

\newtheorem{problem}{Problem}



\usepackage{stackengine}

\newcommand{\m}{\boldsymbol}
\allowdisplaybreaks[4]
\usepackage[colorlinks = true,
linkcolor = blue,
urlcolor  = blue,
citecolor = blue,
anchorcolor = blue]{hyperref}

\usepackage[framemethod=TikZ]{mdframed}
\mdfdefinestyle{MyFrame}{%
	linecolor=black,
	outerlinewidth=1.25pt,
	roundcorner=1.25pt,
	innerrightmargin=5pt,
	innerleftmargin=5pt,}
	
\usepackage[noabbrev]{cleveref}

\usepackage{mathtools}

\DeclarePairedDelimiter\abs{\lvert}{\rvert}%
\DeclarePairedDelimiter\norm{\lVert}{\rVert}%

\makeatletter
\let\oldabs\abs
\def\abs{\@ifstar{\oldabs}{\oldabs*}}
\let\oldnorm\norm
\def\norm{\@ifstar{\oldnorm}{\oldnorm*}}
\makeatother

\usepackage[english]{babel}
\usepackage[utf8]{inputenc}
\usepackage[super]{nth}

\usepackage{graphicx}
\usepackage{float}

\usepackage{array}
\usepackage{threeparttable}

\usepackage[english]{babel}
\usepackage[utf8]{inputenc}
\usepackage[super]{nth}

\RequirePackage{filecontents}

\SetKwRepeat{Do}{do}{while}%

\usepackage{mathtools}

\DeclarePairedDelimiter\evaluat{.}{\rvert}
\reDeclarePairedDelimiterInnerWrapper\evaluat{nostarnonscaled}{%
	\mathopen{}#2\mathclose{#3}%
}
\reDeclarePairedDelimiterInnerWrapper\evaluat{star}{%
	\mathopen{}\mathclose\bgroup #1\hskip -\nulldelimiterspace \relax
	#2\aftergroup\egroup #3%
}

\title{CAV Traffic Control to Mitigate the Impact of Congestion from Bottlenecks: \\ A Linear Quadratic Regulator Approach and Microsimulation Study}

\author{Suyash C. Vishno$\text{i}^{\dagger,\ddagger}$, Junyi J$\text{i}^{\dagger\dagger}$, MirSaleh Bahavarni$\text{a}^{\dagger\dagger}$, Yuhang Zhan$\text{g}^{\dagger\dagger}$, Ahmad F. Tah$\text{a}^{\dagger\dagger}$,\\ Christian G. Claude$\text{l}^{\dagger}$, and Daniel B. Wor$\text{k}^{\dagger\dagger}$
\vspace{-0.5cm}
\thanks{$^\dagger$Department of Civil, Architectural, and Environmental Engineering, The University of Texas at Austin, 301 E. Dean Keeton St. Stop C1700, Austin, TX 78712.
$^{\dagger\dagger}$Department of  Civil and Environmental Engineering, Vanderbilt University, 2201 West End Ave, Nashville, TN 37235.
$^\ddagger$Corresponding author. 
Emails: scvishnoi@utexas.edu, junyi.ji@vanderbilt.edu, mirsaleh.bahavarnia@vanderbilt.edu, yuhang.zhang.1@vanderbilt.edu, 
ahmad.taha@vanderbilt.edu, christian.claudel@utexas.edu, and dan.work@vanderbilt.edu. This work is partially supported
by the National Science Foundation (NSF) under Grants 1636154, 1728629,
1739964, 1917056, 2135579, 2152928, 2152450, and USDOT CAMMSE.
}
}

\newcount\Comments  
\usepackage{color}
\newcommand{\kibitz}[2]{\ifnum\Comments=0\textcolor{#1}{#2}\fi}

\usepackage[noend]{algpseudocode}
\usepackage[center]{caption}
\begin{document}

\maketitle

\begin{abstract}
This work investigates traffic control via controlled connected and automated vehicles (CAVs) using novel controllers derived from the linear-quadratic regulator (LQR) theory. CAV-platoons are modeled as moving bottlenecks impacting the surrounding traffic with their speeds as control inputs. An iterative controller algorithm based on the LQR theory is proposed along with a variant that allows for penalizing abrupt changes in platoons speeds. The controllers use the Lighthill-Whitham-Richards (LWR) model implemented using an extended cell transmission model (CTM) which considers the capacity drop phenomenon for a realistic representation of traffic in congestion. The impact of various parameters of the proposed controller on the control performance is analyzed. The effectiveness of the proposed traffic control algorithms is tested using a traffic control example and compared with existing proportional integral (PI) and model predictive control (MPC) controllers from the literature. A case study using the TransModeler traffic microsimulation software is conducted to test the usability of the proposed controller as well as existing controllers in a realistic setting and derive qualitative insights. It is observed that the proposed controller works well in both settings to mitigate the impact of the jam caused by a fixed bottleneck. The computation time required by the controller is also small making it suitable for real-time control.

\end{abstract}


\section{Introduction}\label{sec:Introduction}
The advent of Connected and Autonomous Vehicle (CAV) technology has led to the opening of unforeseen avenues in the field of traffic control~\cite{delle2019autonomous}. Previously, control was restricted to using actuators that were fixed in space such as variable message signs \cite{liu2015optimize}, or boundary flow controllers \cite{goatin2016speed}. Compared to that, control using CAVs offers greater flexibility as it allows actuators to move in space in a desired manner, therefore, allowing them to be present at desired locations at desired times. In addition to that, using CAVs is relatively cheaper than using fixed actuators which need to be specifically deployed only for the single purpose of traffic control. CAVs on the other hand can be used for several applications like sensing or avoiding hazards due to dangerous driving behavior in their surrounding traffic \cite{yan2018improving}. Also, it can be sometimes difficult to enforce control through traditional fixed actuators like speed limit signs as they can face the issue of low compliance from drivers in some communities. This can also be avoided with the use of CAVs in the traffic stream whose physical presence ahead of drivers would make it impossible for them to avoid the control.

Given the advantages mentioned above, it is essential to explore this newfound potential of traffic control via CAVs by developing new control methodologies that treat CAVs as moving actuators. In this work, we consider the problem of maximizing the mean speed of traffic through traffic jam dissipation by controlling the speed of CAV-platoons entering the road stretch at predefined time intervals using a Linear-Quadratic Regulator (LQR) methodology. CAV-platoons are treated as rolling roadblocks that block the entire flow of traffic at their location. While the general problem of traffic control by controlling the speed of CAV-platoons has been previously explored, the main focus of this work is on proposing and investigating a new controller implementation for this problem in the LQR framework which so far has not been explored in the literature and further compare it with existing approaches from the literature in terms of performance and computational tractability for real-time control.

Several studies in the past decade have considered the problem of moving-bottleneck control of traffic to improve traffic flow. Here a moving-bottleneck implies a reduced flow area that moves along the highway stretch such as that created by slow-moving vehicles. Traditionally, moving bottlenecks are assumed to only partially block the highway cross-section thus allowing part of the traffic to pass by. Unlike that, in this work, the CAV-platoons are considered to block the entire flow of traffic at their location. In \cite{piacentini2019multiple}, the authors have proposed a Proportional-Integral (PI)-type feedback regulator to perform traffic control by controlling the speed of CAV-platoons arriving on the considered highway stretch. While PI-based controllers can produce the desired improvements in traffic flow when coupled with certain arbitrary constraints on the vehicle speeds, in general, they do not guarantee optimal control, and as shown in this study can also result in undesirable control if specific arbitrary bounds on the controlled speeds are removed. In \cite{piacentini2019highway}, the authors propose a model predictive control (MPC)-based speed control algorithm to control the traffic via CAV-platoons subject to the travel time reduction. They solve a nonlinear optimization problem by means of the interior-point algorithm \cite{mehrotra1992implementation} implemented in MATLAB. It considers an extended version of the first-order traffic dynamics \cite{lighthill1955kinematic,richards1956shock} considering the capacity drop phenomenon. Their proposed speed control algorithm is optimal and works well with longer prediction horizon lengths. However, solving a nonlinear optimization problem at each time step is highly time-consuming, especially for extensive networks with several links and junctions, as it requires performing the simulation several times, and therefore can be infeasible for real-time control. In both \cite{piacentini2019highway, piacentini2019multiple}, CAV-platoons are assumed to block the entire flow of traffic at their location. An approach for controlling the speed of the moving-bottlenecks to reduce the overall fuel consumption of the traffic stream is presented in \cite{ramadan2017traffic} which utilizes the wavefront tracking approach to model the interaction between the controlled vehicles and the surrounding traffic described by a first-order traffic model. In comparison to the current study, \cite{ramadan2017traffic} does not consider the capacity drop phenomenon, and also the fuel-consumption-based control approach cannot be extended for traffic flow improvements. \\Besides these articles which deal explicitly with the control aspect of moving-bottlenecks in traffic streams, there are also studies that dive deeper into the accurate modeling of traffic flow dynamics in the presence of moving-bottlenecks at the macroscopic level such as \cite{liard2021entropic, goatin2021interacting, li2020network, villa2017moving, vcivcic2018traffic}.
Note that here we are only interested in studies that use CAV-platoons as moving-bottlenecks. Readers are referred to \cite{delle2019autonomous} for an extensive review of various other use cases associated with CAVs in the realm of traffic control. 

In this work, we utilize the traffic model presented in \cite{piacentini2019highway} which incorporates the capacity drop phenomenon as it allows for realistic control. The authors in \cite{piacentini2019highway} present an MPC-based controller to address the problem of moving-bottleneck control of traffic using CAVs. To overcome the time requirement issue of the MPC-based control algorithm and to make a balance between the quality of the speed control algorithm and its computational requirements, we formulate the traffic control problem in the form of an LQR-based optimization problem which regulates the states around an equilibrium point while utilizing the structure of the state-space dynamics of the system. To solve the LQR-optimization problem, we use the Gauss-Newton $\textrm{LQR}$ (GN-LQR) algorithm which has a time-varying structure since we have to deal with the nonlinearity of the traffic dynamics model via a linear time-varying (LTV) system obtained from the linearization process. Due to the complicated structure of the nonlinearity of the traffic dynamics model corresponding to certain states, we cannot utilize the classic analytic/symbolic methodology to calculate the Jacobian-based state-space matrices of the linearization process. Thus, in those cases, we utilize a numerical methodology developed by \cite{d2006adaptive}, to numerically calculate the Jacobian-based state-space matrices of the linearization process. 

The standard $\textrm{LQR}$ approach and its variants have been used for different control problems in traffic engineering for instance in  \cite{anderson1990optimal,diakaki2002multivariable,wang2022optimizing}. However, in the context of the present traffic control problem using CAV-platoons as moving-bottlenecks, the aforementioned version of LQR is novel.

Besides this, the traffic control studies that address the moving-bottleneck-based control of traffic such as~\cite{piacentini2019multiple,piacentini2019highway,ramadan2017traffic} have only been carried out using macroscopic traffic simulations. While macroscopic traffic models are attractive due to their robustness and scalability, they are not always realistic which imposes questions on whether such controllers which use macroscopic traffic models at their core are useful in the real-world setting. To address this gap, we also present a microscopic traffic simulation-based case study that tests the proposed LQR-based controller under realistic settings and tries to address questions about the usability and corresponding gaps in the application of such controllers in the real world. 

Given the main research gap in this area is the absence of an optimal controller offering fast computation of controls for real-time moving-bottleneck control of traffic and the absence of a study on the moving-bottleneck controllers under realistic settings, the present study makes the following contributions:
\begin{enumerate}
\item An LQR-based controller design with macroscopic model dynamics is proposed to control the speeds of CAV-platoons allowing for mitigation of the effect of jam-forming bottlenecks in the traffic stream. The LQR-based controller uses the structure of the state-space matrices of the traffic dynamics system and does not require performing repeated simulations for control, therefore requiring less computation time. The impact of various parameters of the LQR-based controller is investigated with respect to its performance in solving the given problem.
\item A variant of the LQR-based controller allowing for a penalty on large changes in control inputs over consecutive time steps is developed and shown to reduce the magnitude of fluctuations in the controlled speeds allowing for safe and realistic control.
\item We present a comparison of the proposed LQR-based controllers with existing MPC-based \cite{piacentini2019highway} and PI-based \cite{piacentini2019multiple} controllers from the literature in terms of computational tractability and performance using macroscopic simulation. The proposed LQR-based controllers are observed to perform similarly to the existing controllers in terms of improvement in traffic conditions and outperform them in terms of computation time by about two orders of magnitude in the given traffic scenario (with PI-based controllers this is true when the controller requires tuning the controller gains in real-time).
\item The performance of the proposed LQR-based controller is further investigated using a microscopic traffic simulation setup and compared with the application of existing controllers to the same setup to assess its applicability and utility under realistic settings of traffic flow.

\end{enumerate}
The remainder of the article is organized as follows- Section \ref{s:traffic_dynamics_model} describes the traffic dynamics model used in this work. The problem statement with the LQR-based solution scheme and algorithms is presented in Section \ref{s:problem_statement}. Section \ref{s:numerical_study} proposes research questions related to the problem, analyzes the proposed approach in a macroscopic setting, and compares it with existing approaches from the literature, followed by a microsimulation-based case study on a similar setup using the  proposed and existing controllers. The article is concluded with Section \ref{s:conclusion} which presents preliminary answers to the proposed research questions and proposes directions for future work.

\textbf{Notations:} We denote the vectors and matrices by lowercase and uppercase bold symbols, respectively. The set of $m$-dimensional real-valued vectors and $n \times p$ real-valued matrices are denoted by $\mathbb{R}^m$ and $\mathbb{R}^{n \times p}$, respectively. The identity matrix of dimension $q$ is represented by $\m I_q$. The vector/matrix transpose is denoted by $^T$. The positive semi-definiteness and positive definiteness are represented by $\succeq \m 0$ and $\succ \m 0$, respectively. The set-theoretical minimum operator is denoted by $\min$. The dependency on the discrete-time time index $k$ is shown by $[k]$. The prefix $\delta$ adding to any time-varying quantity represents the linearized LTV dynamics value, i.e., the difference between the nonlinear dynamics value and the corresponding equilibrium value.

\section{Traffic Dynamics Model} \label{s:traffic_dynamics_model}
Here, we present the state-space formulation for the traffic dynamics model considered in this work. The flow of traffic across a highway stretch with no on-ramps or off-ramps is modeled using the first-order LWR model \cite{lighthill1955kinematic,richards1956shock} while accounting for the capacity drop phenomenon \cite{yuan2015capacity,kontorinaki2017first}. The model is implemented using a Godunov scheme \cite{Godunov1959} which is proposed previously in \cite{piacentini2019highway,han2017resolving} and is an extension of the classical Cell Transmission Model (CTM) implementation proposed in \cite{Daganzo1994}. Within this, the highway stretch is divided into $N_L$ segments of equal length $L$ (km) and the time horizon is divided into $N_T$ smaller duration of $T$ (sec) each such that the Courant-Friedrichs-Lewy (CFL) condition~\cite{courant1967partial}: $T\le L/v_f$ is satisfied where $v_f$ refers to the free-flow speed of traffic. Let $N_{\mathrm{CAV}}$ be the total number of controlled CAV-platoons currently on the modeled highway stretch. The traffic dynamics model is given as follows:
\begin{align*}
    \rho_i[k+1] &= \rho_i[k] + (T/L)(\phi_i(\rho_i[k], {\m u}[k])-\phi_{i+1}(\rho_i[k],{\m u}[k]),
\end{align*}
$\forall i\in\{1,\dots,N_L\}$, where $\rho_i[k]$ represents the traffic density (vehicles per unit length) in Segment $i$ at time index $k$, $\m u[k] \in \mathbb{R}^{N_{\mathrm{CAV}}}$ denotes the control input and is given as
\begin{align}
\label{e:input_vector}
    & {\m u}[k]=[{u}_1[k]\dots  {u}_{N_{\mathrm{CAV}}}[k]]^T,
\end{align} 
where ${u}_j[k], \forall j\in\{1,\dots,N_{\mathrm{CAV}}\}$ denotes the control speed of CAV-platoon $j$ in the traffic stream. $\phi_i(. , .)$ is the actual traffic flow (vehicles per unit time) that leaves Segment $i$ and is given as 
\begin{align}
    &\hspace{-0.1cm} \phi_i(\rho_i[k],{u}_j[k]) = \min\{D_i(\rho_i[k],{u}_j[k]),S_{i+1}(\rho_{i+1}[k])\}, \label{e:flow}
\end{align}
assuming the CAV-platoon $j$ is in Segment $i$ at time index $k$. The demand and supply functions are further defined using minimum functions of the state and input variables respectively as follows:
\begin{align*}
    D_i(\rho_i[k],{u}_j[k]) &= \min\{v_i({u}_j[k])\rho_i[k],q^{max}_{i}(\rho_i[k])\},\quad
    S_i(\rho_i[k]) = w_c(\rho_m-\rho_i[k]),
\end{align*}
where
\begin{align*}
    q^{max}_i(\rho_i[k]) &= q_i^{cap} \times \min \bigg \{1,1+(\alpha-1)\dfrac{\rho_i[k]-\rho_c}{\rho_m-\rho_c}\bigg \}.
\end{align*}
Here, $q_i^{cap}$ denotes the maximum capacity of Segment $i$, $\rho_c, \rho_m, w_c$ are parameters of the triangular fundamental diagram of traffic flow denoting the critical density, the maximum density, and the maximum congestion wave speed of traffic, respectively, and $\alpha \in [0,1]$ is a coefficient denoting the extent of the capacity drop where $\alpha=1$ implies no capacity drop. Here, $v_i({u}_j[k])$ denotes the maximum speed of traffic in Segment $i$ at time index $k$ which is given by the following conditional:
\begin{align*}
    v_i({u}_j[k])&=
    \begin{cases}
    {u}_j[k], & \mathrm{if~CAV~platoon}~j~\mathrm{is~in~Segment}~ i~\mathrm{at}~k,\\
    v_f, & \mathrm{otherwise}.
    \end{cases}
\end{align*}
It is noteworthy this is only the maximum possible speed of traffic in Segment $i$ and not necessarily the actual speed since the actual speed will depend on the realized flow which is a function of both the demand and the supply as shown in \eqref{e:flow}. In the sequel, for convenience, we denote the demand, supply, and actual flow with the function names followed by the time index without mentioning the inputs required to calculate each.
\begin{figure}[!ht]
    \centering
    \includegraphics[width=0.5\textwidth]{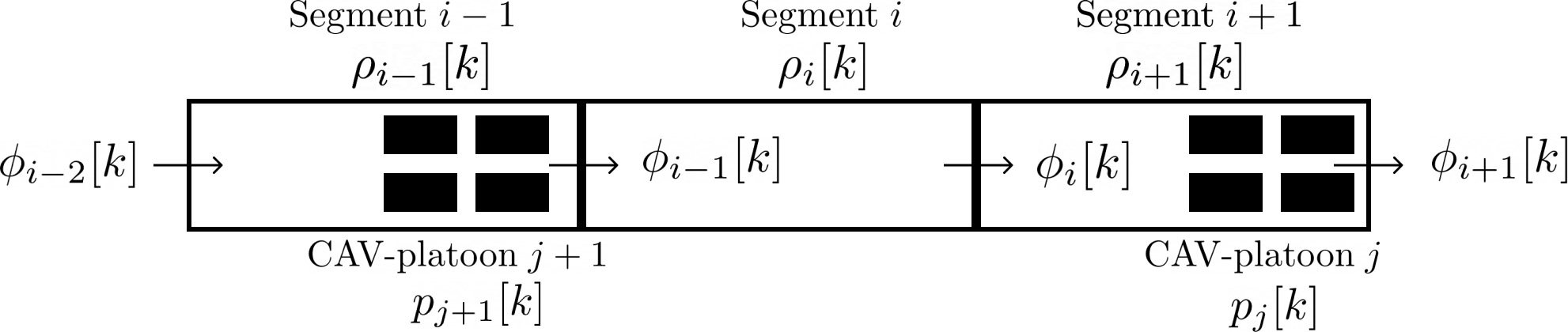}
    \caption{Three segments of the modeled highway stretch along with two CAV-platoons and the corresponding states written underneath. Arrows indicate the direction of traffic flow.}
    \label{f:schematic}
\end{figure}

The position of CAV-platoon $j$ on the highway is denoted by $p_j[k]$ where its evolution over time is given as 
\begin{align}
\label{e:trajectory_update}
    p_j[k+1]&=p_j[k]+T\bar{v}_j(\rho_i[k],\rho_{i+1}[k],{u}_j[k]),
\end{align}
where $\bar{v}_j(\rho_i[k],\rho_{i+1}[k],{u}_j[k])$ denotes the speed of CAV-platoon $j$ during time index $k$. Note that $u_j[k]$ is the control speed of the CAV-platoon or the speed prescribed to the CAV-platoon by the controller while $\bar{v}_j$ is the realized speed of the CAV-platoon which depends on the demand and supply conditions besides the control speed.

Here the bar on top of $v$ is used to differentiate the speed of the CAV from the maximum speed of a segment which is also denoted by $v$. Figure \ref{f:schematic} presents a schematic of the highway stretch with the two elements-Segments and CAV-platoons along with their associated states written underneath each label. 

In the following, we also denote this final speed by the function name followed by the time index with the CAV-platoon-index as a subscript. If CAV $j$ is in Segment $i$ at time index $k$ and is expected to end up in Segment $i$ at the end of this time step while traveling at its control speed or if $D_i[k]\le S_{i+1}[k]$, that is there is no restriction on the flow wanting to leave Segment $i$ at time $k$, then its final speed can be set directly as $\bar{v}_j[k]={u}_j[k]$ and its final position is calculated using \eqref{e:trajectory_update}. 
On the other hand, if CAV-platoon $j$ is expected to end up in Segment $i+1$ at the control speed and $D_i[k]>S_{i+1}[k]$ that is the flow is restricted by the downstream segment, then its final speed and hence final position needs to be calculated according to certain conditions which are presented in detail in \cite{piacentini2019multiple} and depend on the platoon length denoted by $l_j$ (m), and the minimum demand needed for the platoon to pass to the next segment denoted by $S_{min}$ which is assigned an arbitrary value,

apart from the variables and functions introduced above. Thus, \eqref{e:trajectory_update} is, in fact, nonlinear as the calculation of $\bar{v}_j[k]$ requires the evaluation of conditional statements.

The state-space equation can therefore be written as
\begin{empheq}[box=\fbox]{align}
    {\m x}[k+1] &= {\m A} {\m x}[k] + {\m G} {\m f}({\m x}[k],{\m u}[k]), \label{SSM}
\end{empheq}
where the state vector at time index $k$
\begin{align*}
    {\m x}[k]&=[\rho_1[k] \dots \rho_{N_L}[k] \hspace{2mm} p_1[k] \dots p_{N_{\mathrm{CAV}}}[k]]^T\in \mathbb{R}^{N_L+N_{\mathrm{CAV}}},
\end{align*}
consists of the traffic densities from all the segments and the current positions of the CAV-platoons on the highway, and the input vector is the same as in \eqref{e:input_vector} consisting of the control speeds for all the CAV-platoons on the highway stretch. 

Let $n_x := N_L+N_{\mathrm{CAV}}$ be the number of states and $n_u := N_{\mathrm{CAV}}$ be the number of inputs. The matrix ${\m A}= {\m I_{n_x}}$, matrix ${\m G} \in\mathbb{R}^{n_x\times n_x}$ is a diagonal matrix representing the coefficients of the nonlinearities in the dynamics as
\begin{align*}
    {\m G} &= T \begin{bmatrix}
    {\m I_{N_L}}/L & {\m 0}\\
    {\m 0} & {\m I}_{N_{\mathrm{CAV}}}
    \end{bmatrix},
\end{align*}
and the vector-valued function ${\m f}:\mathbb{R}^{n_x} \times \mathbb{R}^{n_u} \rightarrow \mathbb{R}^{n_x}$ represents the nonlinearities in the evolution of traffic density and the position of the CAVs with time. In particular, the vector ${\m f}({\m x}[k],{\m u}[k])$ can be written as 
\small
\begin{align}
\label{e:nonlinear_state_space}
    {\m f}({\m x}[k],{\m u}[k]) &= \begin{bmatrix}
    \phi_0({\m x}[k], {\m u}[k])-\phi_1( {\m x}[k], {\m u}[k]) \\
    \vdots \\
    \phi_{N_L-1}({\m x}[k], {\m u}[k])-\phi_{N_L}({\m x}[k], {\m u}[k]) \\
    \bar{v}_1({\m x}[k], {\m u}[k]) \\
    \vdots \\
    \bar{v}_{N_{\mathrm{CAV}}}({\m x}[k], {\m u}[k])    
    \end{bmatrix}.
\end{align}
\normalsize
The nonlinearity is indeed non-trivial since it consists of differences of nested minimum functions \eqref{e:flow}
as well as CAV-platoon speeds obtained from nested conditional statements. The presence of such nonlinearity in the state space makes it necessary for control problems based on the model to utilize nonlinear optimization schemes.
 
In the next section, we formally define the traffic control problem considered in this study along with the control methodology used to address it.
\vspace{-2mm}
\section{Problem Statement and \textrm{LQR}-Based Traffic Control Algorithms} \label{s:problem_statement}
The underlying traffic control problem addressed in this work is defined as follows:

\begin{problem} \label{Pro1}
Given the nonlinear traffic dynamics \eqref{SSM}, control the speed of CAV-platoons entering the highway stretch at known time steps to mitigate the adverse effects of a traffic jam formed in the middle of the stretch.
\end{problem}
\vspace{1cm}
 Problem \ref{Pro1} can be defined in the form of an optimization problem as follows:
\begin{align}
    \min_{u[k]} \quad & J(\m x[k], \m u[k]) \nonumber \\
    \mathrm{s.t.} \quad & \eqref{SSM} \nonumber \\
    \quad & \m u[k] \in \mathcal{U} 
    \label{e:optimization1}
\end{align}
where the cost function $J(\m x[k], \m u[k])$ is any function whose minimization ensures an improvement in the traffic conditions which can be in terms of an increase in the overall speed of traffic or a decrease in the overall congestion level on the highway in terms of traffic density. Here, the decision variables $\m u[k]$ are the speeds of the CAV-platoons on the highway stretch. The essential constraints include the state-space dynamics \eqref{SSM} while the speeds of these platoons can also be constrained to an arbitrary set $\mathcal{U}$.

In the present work, the optimization problem \eqref{e:optimization1} is formulated in the LQR optimization framework~\cite{boyd2008linear}. For linear systems, this results in a horizon-based optimization problem that aims to regulate the states and inputs of the system around the zero point taking into account the system dynamics over a given number of future time steps with the help of a state-feedback law for the control input in the form $\m u[k]=\m K[k]\m x[k]$ where $\m K[k]$ is called the gain matrix and is calculated using existing formulae from the literature. For nonlinear systems, an LQR-based optimization problem can be written by linearizing the system around an equilibrium point over the length of the horizon and regulating the difference between the actual state (respectively input) and the equilibrium state (respectively input) around the zero point which results in the control input trying to bring the system closer to the equilibrium states. In the context of traffic control, these equilibrium states and inputs are assigned values that result in an improvement in the state of traffic. In this case, the control input is defined by the following state-feedback law which takes into account the selected equilibrium states and inputs:
\begin{empheq}[box=\fbox]{align}
    {\m u}[k] &= -{\m K}[k]{\m x}[k] + {\m u^{\m \ast}}[k] + {\m K}[k] {\m x^{\m \ast}}[k], \label{uKke}
\end{empheq}
where ${\m K}[k] \in \mathbb{R}^{n_u \times n_x}$ and $({\m x^{\m \ast}}[k],{\m u^{\m \ast}}[k])$ denote the time-varying $\textrm{LQR}$ state-feedback matrix and the time-varying equilibrium point of the nonlinear system \eqref{SSM} at time index $k$, respectively. 

To obtain the gain matrix $\m K[k]$ at any time-step for controlling the nonlinear system within the LQR framework, the Gauss-Newton $\textrm{LQR}$ algorithm \cite{boyd2008linear} can be applied. The same is presented in the remainder of this section along with a variant of the GN-LQR algorithm that penalizes changes in control inputs over consecutive time steps. Various parameters of these algorithms are investigated in the ensuing sections in the context of traffic control using moving-bottlenecks.

\subsection{The Gauss-Newton \textrm{LQR} algorithm}\label{GNLQR}
Here, we present an iterative $\textrm{LQR}$ algorithm called the GN-LQR algorithm \cite{boyd2008linear} which can be used to solve the LQR optimization problem \eqref{e:optimization1} for the given nonlinear system \eqref{SSM}. We introduce the following notation before presenting the GN-LQR algorithm:\\
$N$: \textrm{horizon length}.\\
$N$-step input and state matrices:
\begin{align*}
{\m U} &:= \begin{bmatrix} {\m u}[0] & \dots & {\m u}[N-1] \end{bmatrix},\quad
{\m X} := \begin{bmatrix} {\m x}[0] & \dots & {\m x}[N] \end{bmatrix}.
\end{align*}
Corresponding time-varying equilibrium counterparts:
\begin{align*}
{\m U^{\ast}} &:= \begin{bmatrix} {\m u^{\ast}}[0] & \dots & {\m u^{\ast}}[N-1] \end{bmatrix},\quad
{\m X^{\ast}} := \begin{bmatrix} {\m x^{\ast}}[0] & \dots & {\m x^{\ast}}[N] \end{bmatrix}.
\end{align*}
Corresponding control input difference  matrix:
\begin{align*}
{\m \delta \m U} &:= \begin{bmatrix} {\m \delta \m u}[0] & \dots & {\m \delta \m u}[N-1] \end{bmatrix},
\end{align*}
Linearized state-space matrices:
\begin{align}\label{e:linearized_state_space}
    \m{\hat{A}}[k] &= \m{A} + \m{G} \m{A_f}[k],\quad
    \m{\hat{B}}[k] = \m{G} \m{B_f}[k],
\end{align}
where\\
$\m{A}_f[k]$ : The derivative matrix of $\m{f}(\m{x}[k],\m{u}[k])$ w.r.t. $\m x^{\m \ast}[k]$,\quad
$\m{B_f}[k]$ : The derivative matrix of $\m{f}(\m{x}[k],\m{u}[k])$ w.r.t. $\m u^{\m \ast}[k]$.\\
The LQR cost function to be minimized:
\begin{align}
\label{e:LQR_objective}
J(\m{x}[k],\m{u}[k]) &:= J_x(\m{x}[k]) + J_u(\m{u}[k]),
\end{align}
where
\begin{align*}
    J_x(\m{x}[k]) &:= \sum_{k = 0}^{N-1} (\m{x^{\ast}}[k] + \m{\delta x}[k])^T \m{Q} (\m{x^{\ast}}[k] + \m{\delta x}[k]),\quad
    \m{Q} : \textrm{The LQR state-weight matrix}, \m{Q} \succeq \m{0},\\
    J_u(\m{u}[k]) &:= \sum_{k = 0}^{N-1} (\m{u^{\ast}}[k] + \m{\delta u}[k])^T \m{R} (\m{x^{\ast}}[k] + \m{\delta u}[k]),\quad
    \m{R} : \textrm{The $\textrm{LQR}$ input-weight matrix}, \m{R} \succ \m{0}.
\end{align*}
The goal of the algorithm is to minimize the above objective function given the state-space dynamics \eqref{SSM} along with physical bounds on the speeds. With the above notation, the Gauss-Newton $\textrm{LQR}$ ($\textrm{GN-LQR}$) algorithm \cite{boyd2008linear} can be summarized in Algorithm \ref{alg:cap}. To the standard algorithm, we also add a step to impose a non-negativity constraint and an upper bound on the speed equal to the free-flow speed.

\setlength{\floatsep}{5pt}{
\begin{algorithm}[t]
\caption{\textbf{The $\textrm{GN-LQR}$ Algorithm}}\label{alg:cap}
\DontPrintSemicolon
\textbf{input:} State-space matrices $\m{A}$, $\m{G}$, nonlinear function $\m{f}$, initial state $\m{x}[0]$, horizon length $N$, LQR weight matrices $\m{Q}$, $\m{R}$, error tolerance $\epsilon$, maximum number of iterations $M$, initial guess for equilibrium control inputs ${\m U}^{\m \ast}$, and initial guess for initial equilibrium state ${\m x}^{\m \ast}[0]$.

\textbf{set:} current iterate $i=0$, $\m U=\m U^{\m \ast}$, $\m{\delta U}=\m U-\m U^{\m \ast}$, $\m{\delta x}[0] = \m{x}[0] - \m{x^{\ast}}[0]$.

\Repeat{$\|\m{\delta U}\| < \epsilon$ $\textrm{or}$ $i > M$}{
\For{$k = 0,\dots,N-1$}{
\textbf{compute:} $\hat{\m A}[k], \hat{\m B}[k]$ via \eqref{e:linearized_state_space} around the time-varying equilibrium point $({\m x^{\m \ast}}[k],{\m u^{\m \ast}}[k])$ of nonlinear dynamics \eqref{SSM} at time index $k$. 

\textbf{set:} $\m{\delta x}[k+1] = \m{\hat{A}}[k] \m{\delta x}[k] + \m{\hat{B}}[k] \m{\delta u}[k]$. 

\textbf{compute:} ${\m x}^{\m \ast}[k+1]$ via nonlinear dynamics \eqref{SSM}.
}
\Comment{Solve the Gauss-Newton optimization problem for controller gains $\m{K}[0], \dots, \m{K}[N-1]$}

\textbf{set:} $\m{P}[N] = \m{0}$ 

\For{$l = N,\dots,1$}{
\textbf{set:}\hspace{2mm}$\m{P}[l-1]  = \m{Q} + \m{\hat{A}}[l-1]^T \m{P}[l] \m{\hat{A}}[l-1] -\m{\hat{A}}[l-1]^T \m{P}[l] \m{\hat{B}}[l-1] \times (\m{R} + \m{\hat{B}}[l-1]^T \m{P}[l] \m{\hat{B}}[l-1])^{-1}\m{\hat{B}}[l-1]^T \m{P}[l] \m{\hat{A}}[l-1]$.}
\For{$k = 0,\dots,N-1$}{
\textbf{set:} $\m{K}[k] = (\m{R} + \m{\hat{B}}[k]^T \m{P}[k+1] \m{\hat{B}}[k])^{-1} \m{\hat{B}}[k]^T \m{P}[k+1] \m{\hat{A}}[k]$.

\textbf{set:} $\m{\delta u}[k] = -\m{K}[k] \m{\delta x}[k]$.
}
 
\textbf{set:} $\m{\delta U}=\begin{bmatrix} {\m \delta \m u}[0] & \dots & {\m \delta \m u}[N-1] \end{bmatrix}$

\textbf{set:} $\m{U} = \min\{\max\{\m{U^{\ast}} + \m{\delta U},0\},v_f\}$, $\m{U^{\ast}} = \m{U}$, $i = i+1$.
}

\textbf{compute:} $\m u[0]$ via \eqref{uKke} using $\m{K}[0]$.

\textbf{output:} $\m u[0]$.
\end{algorithm}
}

\subsection{The Gauss-Newton LQR algorithm with a penalty on variation in inputs}\label{GNLQRP}
The controls produced at any time step using the GN-LQR controller are independent of the controls in the previous time steps. Due to this, the optimal controls can vary significantly over consecutive time steps as is observed in Section \ref{s:numerical_study}. Since these controls are executed by CAV-platoons that are traveling within a traffic stream comprised of both autonomous and human-driven vehicles, the latter of which can sometimes have high reaction times, large changes in control inputs over consecutive time steps can result in life-threatening collisions due to vehicles not braking in time. To avoid such circumstances, here we present a variant of the LQR optimization problem which applies a penalty on changes in control inputs over consecutive time steps thus preventing large changes in control inputs. The implementation of the optimization problem is derived based on \cite{pieterabbeel} which prescribes the inclusion of an additional term in the LQR objective function penalizing large variations in control inputs.

This is achieved by modifying the state-space formulation of the system by defining a new state which is an augmentation of the original state vector and the original control input vector and a new control input vector that captures the change in control input. For linear systems, the derivation of the new augmented system and a new LQR objective is provided in Appendix \ref{a:GN-LQRP}. A new weight matrix $\m R'$ is introduced in the LQR optimization problem that governs the fluctuations in the control inputs. A larger magnitude of elements in $\m R'$ implies a larger penalty on the change in control inputs over consecutive time steps whereas $\m R' = \m 0$ implies no penalty is imposed and the resulting optimization is equivalent to the standard LQR optimization problem. The GN-LQR algorithm presented in the previous section is modified in Step 4 to obtain the new algorithm, Algorithm \ref{alg:cap2}, which is referred to as GN-LQR-with-penalty (GN-LQRP) in the remainder of the article.

\setlength{\floatsep}{5pt}{
\begin{algorithm}[t]
\caption{\textbf{The $\textrm{GN-LQRP}$ Algorithm}}\label{alg:cap2}
\DontPrintSemicolon
\textbf{input:} State-space matrices $\m{A}$, $\m{G}$, nonlinear function $\m{f}$, initial state $\m{x}[0]$, horizon length $N$, LQR weight matrices $\m{Q}$, $\m{R}$, $\m{R}'$, error tolerance $\epsilon$, maximum number of iterations $M$, initial guess for equilibrium control inputs ${\m U}^{\m \ast}$, and initial guess for initial equilibrium state ${\m x}^{\m \ast}[0]$, previous control input $\m u_o$.

\textbf{set:} $\m U=\m U^{\m \ast}$, $\m{\delta U}=\m U-\m U^{\m \ast}$.

\textbf{compute:} $\m Q'$ via \eqref{e:augQ}, $\m x'[0]$, $\m u'[0]$, $\m {x'}^{\m \ast}[0]$, $\m {u'}^{\m \ast}[0]$ via \eqref{e:augmented_matrices} using $\m u_o$, and $\m\delta\m x'[0]=\m x'[0]-\m {x'}^{\m \ast}[0]$.

\textbf{compute:} $\m U', \m {U'}^{\m \ast}$ using \eqref{e:augmented_matrices}.

\textbf{set:} current iterate $i=0$.

\Repeat{$\|\m{\delta U}\| < \epsilon$ $\textrm{or}$ $i > M$}{
\For{$k = 0,\dots,N-1$}{

\textbf{compute:} $\hat{\m A}[k], \hat{\m B}[k]$ via \eqref{e:linearized_state_space} around the time-varying equilibrium point $({\m x^{\m \ast}}[k],{\m u^{\m \ast}}[k])$ of nonlinear dynamics \eqref{SSM} at time index $k$.

\textbf{set:} $\m{\delta x}[k+1] = \m{\hat{A}}[k] \m{\delta x}[k] + \m{\hat{B}}[k] \m{\delta u}[k]$.

\textbf{compute:} ${\m x}^{\m \ast}[k+1]$ via nonlinear dynamics \eqref{SSM}, and $\m \delta \m x'[k+1]$ via \eqref{e:augmented_matrices}.
}
\Comment{Solve the Gauss-Newton optimization problem with penalty for controller gains $\m{K}[0], \dots, \m{K}[N-1]$}

\textbf{set:} $\m{P}[N] = \m{0}$ 

\For{$l = N,\dots,1$}{
\textbf{compute:} $\hat{\m A}'[l-1]$, $\hat{\m B}'[l-1]$ via \eqref{eq:augmented_state_space}.

\textbf{set:}\hspace{2mm}$\m{P}[l-1]  = \m{Q}' + \m{\hat{A}}'[l-1]^T \m{P}[l] \m{\hat{A}}'[l-1] -\m{\hat{A}}'[l-1]^T \m{P}[l] \m{\hat{B}}'[l-1] \times (\m{R}' + \m{\hat{B}}'[l-1]^T \m{P}[l] \m{\hat{B}}'[l-1])^{-1}\m{\hat{B}}'[l-1]^T \m{P}[l] \m{\hat{A}}'[l-1]$.}
\For{$k = 0,\dots,N-1$}{
\textbf{set:} $\m{K}[k] = (\m{R}' + \m{\hat{B}}'[k]^T \m{P}[k+1] \m{\hat{B}}'[k])^{-1} \m{\hat{B}}'[k]^T \m{P}[k+1] \m{\hat{A}}'[k]$.

\textbf{set:} $\m{\delta u}'[k] = -\m{K}[k] \m{\delta x}'[k]$.
}
 
\textbf{set:} $\m{\delta U}'=\begin{bmatrix} {\m \delta \m u}'[0] & \dots & {\m \delta \m u}'[N-1] \end{bmatrix}$

\textbf{set:} $\m{U}' = \m{U'}^{\m\ast}+\m{\delta U}'$.

\For{$k=N-1,\dots,1$}{
\textbf{set:} $\m u[k] = \m u[k-1] + \m u'[k]$
}

\textbf{set:} $\m u[0] = \m u_o + \m u'[0]$

\textbf{set:} $\m U=\begin{bmatrix} {\m u}[0] & \dots & {\m u}[N-1] \end{bmatrix}$.

\textbf{set:} $\m U = \min\{\max\{\m U, 0\}, v_f\}$

\textbf{set:} $\m{\delta U}=\m U - \m U^{\m\ast}$

\textbf{set:} $\m{U^{\ast}} = \m{U}$, $i = i+1$.
}
\textbf{output:} $\m u[0]$.

\end{algorithm}
}

In the next section, we present the impact of various parameters of the LQR-based algorithms on the performance of the controller and compare it with existing controllers from the literature.

\section{Numerical Study And Implementation}
\label{s:numerical_study}
In this section, we investigate the performance of the proposed control algorithms for moving-bottleneck-based traffic control, mainly in the mitigation of the impact of traffic jams on a highway stretch. A primary investigation is carried out using macroscopic simulations performed with the CTM model described in Section \ref{s:traffic_dynamics_model} where the best-case performance of the controller is observed, its sensitivity to various parameters of the algorithm is examined, and comparisons are made with existing PI-~\cite{piacentini2019multiple} and MPC-based \cite{piacentini2019highway} controllers. Details of the implementation of the latter two controllers are also presented as part of the analysis. This is followed by a microsimulation-based case study using the Transmodeler traffic simulation software to test the near real-world performance of the controller and to learn the advantages and gaps in applying the controller in the real world. All macroscopic simulations applying the CTM model are performed using MATLAB R2021b running on a $64$-bit Windows $10$ with a $2.2$GHz IntelR CoreTM i$7$-$8750$H CPU and 16GB of RAM.

\subsection{Numerical study objectives}
\label{s:study_objectives}
The goal of this study is to find the answers to the following questions:
\begin{itemize}
    \item \textit{Q1:} Are the LQR-based controllers, namely GN-LQR and GN-LQRP, able to reduce the impact of bottlenecks on the highway traffic flow? How do they perform compared with PI- and MPC-based controllers proposed in the literature?
    \item \textit{Q2:} Are the LQR-based controllers computationally feasible for real-time traffic control? How do they compare with the PI- and MPC-based controllers in terms of computational tractability?
    \item \textit{Q3:} Are the controls produced by the controllers realistic with regard to application in the real world?
    \item \textit{Q4:} What is the impact of the various LQR-based controller parameters on its performance with respect to traffic control? How do the horizon length, number of iterations, and LQR weight parameters impact the control speeds of the CAV-platoons?
    \item \textit{Q5:} How do the controllers perform in a realistic microscopic traffic setting? Is the performance comparable to the macroscopic case?
\end{itemize}
What follows is a description of the traffic flow scenario and evaluation metrics used to test the performance of the controllers and for comparison with existing techniques.

\subsection{Scenario description and evaluation metrics}
\label{s:scenario_and_metrics}
\begin{figure}
    \centering
    \includegraphics[width = 0.4\textwidth]{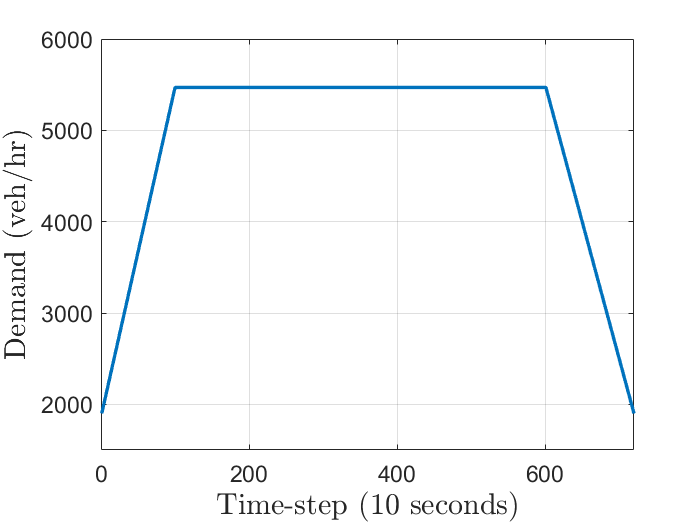}
    \vspace{-2mm}
    \caption{Upstream demand profile for the given example.}
    \label{f:demand_profile}
\end{figure}

The traffic is modeled using the dynamics presented in Section \ref{s:traffic_dynamics_model}. This section presents the values of the traffic parameters introduced in Section \ref{s:traffic_dynamics_model} along with a description of the default traffic scenario without any fixed bottleneck on the highway stretch, the uncontrolled scenario in the presence of a fixed bottleneck, and the evaluation metrics used to quantitatively compare between different scenarios and controllers.  We consider an $8$ km long highway stretch with no on-ramps or off-ramps which is divided into $16$ even segments of length $0.5$ km each. A total duration of $2$ hr is considered for the example with time divided into steps of the duration of $10$ sec each. The following values of traffic flow parameters are considered: $\rho_c=60$ veh/km, $v_f=100$ km/hr, $w_c=38$ km/hr, $\rho_m=320$ veh/km, $q_{max}=6000$ veh/hr, and $\alpha=0.83$, similar to \cite{piacentini2019highway}. We consider a platoon length of $4.5$ m which essentially implies platoons of one CAV per lane, and $S_{min}=10$. The initial density on all the segments is set to $20$ veh/km. The available supply at the downstream end of the highway is set to $q_{max}$ while the demand wanting to enter at the upstream end of the highway has the profile shown in Figure \ref{f:demand_profile} where the starting and ending demand is $1900$ veh/hr and the value along the horizontal line is $5490$ veh/hr. A reduced flow area is simulated on the highway by reducing the outflow of Segment $13$ to $5400$ veh/hr for the first hour after which the flow of the segment is restored to maximum capacity. The impact of the control is measured using three metrics, the Total Travel Time ($\mathrm{TTT}$) in veh$\cdot$hr, the Total Travel Distance ($\mathrm{TTD}$) in veh$\cdot$km, and the Mean Speed ($\mathrm{MS}$) in km/hr which are defined similarly to \cite{piacentini2019highway} as follows:
\begin{align*}
    & \mathrm{TTT} = TL\sum_{k=1}^{N_T}\sum_{i=1}^{N_L}\rho_i[k],\quad \mathrm{TTD} = TL\sum_{k=1}^{N_T}\sum_{i=1}^{N_L}\phi_i[k], \quad \textrm{and  
  }\mathrm{MS}= \mathrm{TTD}/\mathrm{TTT},
\end{align*}
where $T, L, N_T$, and $N_L$ are the duration of each time step, the length of each segment, the total number of time steps in the simulation, and the total number of segments  in the considered highway stretch, respectively. In general, a lower $\mathrm{TTT}$, a higher $\mathrm{TTD}$, and a higher $\mathrm{MS}$ are desirable --- The closer the traffic density is to the critical density, the better the values of these metrics as the traffic is free-flowing and at the maximum flow possible. Therefore, at each implementation of the LQR-based controllers we select such equilibrium states for linearization which improves the values of these metrics. In addition to these metrics, in order to compare the computational performance of different controllers, we consider the Average Computation Time (ACT) for each controller which is defined as the average time required to compute the control inputs at any time step during the simulation. It is computed simply as the average of the time consumed over all the runs of the controller during the simulation.

\begin{figure}
    \centering    \includegraphics[width=0.4\textwidth]{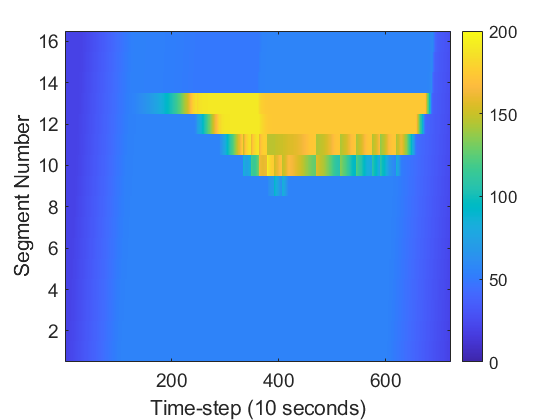}
    \caption{Density (veh/km) evolution in the uncontrolled case with the reduced bottleneck flow.}
    \label{f:no_control}
\end{figure}

We assume that controlled vehicles enter the stretch every $15$ time steps starting from time step $60$ to time step $600$ of the process horizon. Under normal circumstances, that is in the absence of the reduced flow area, $\mathrm{TTT}=790$ and $\mathrm{TTD}=78,998$ which gives $\mathrm{MS}=99.9$. Figure \ref{f:no_control} presents the simulated traffic densities in the presence of the reduced flow and without any control implementation, that is, ${u}_j[k]=v_f, \forall j \in \{1,2,\dots,N_{\mathrm{CAV}}\}$. Note that here, $N_{\mathrm{CAV}}$ is used to refer to the total number of CAV-platoons that will enter the highway stretch during the entire simulation and not the number of CAV-platoons present on the stretch at once which varies with time. In this case, the evaluation metrics are $\mathrm{TTT}=1,019$ and $\mathrm{TTD}=78,998$, and $\mathrm{MS}=77.5$. As expected the reduced flow area creates temporary congestion on the highway which results in longer travel times for the same travel distance and therefore a lower $\textrm{MS}$. In the following sections, we elaborate on the implementation aspects of the LQR-based controllers including the selection of parameters and their impact on the controller's performance followed by a comparison of the proposed controller with existing controllers from the literature. The implementation details of the existing controllers are presented in Section \ref{s:comparison} and in Appendix \ref{a:PI} and Appendix \ref{a:MPC}.

\subsection{Impact of LQR parameters}
Here, we test the impact of various parameters of the LQR-based controllers in the context of traffic control using CAV-platoons as moving-bottlenecks. A default set of parameters is defined for the proposed controllers and different parameters are varied in isolation to assess their impact on the control performance which is measured using the parameters defined in Section \ref{s:scenario_and_metrics}, namely the $\textrm{TTT}$, $\textrm{TTD}$, and $\textrm{MS}$.

\subsubsection{Default LQR-based controller implementation}
\label{s:lqr_default}
In this section, we describe the default parameter values for the two LQR-based controllers, namely the GN-LQR (Algorithm \ref{alg:cap}) and the GN-LQRP (Algorithm \ref{alg:cap2}) controllers. In the following sections, which test the impact of different parameters of the algorithm on their performance, the values of individual parameters are varied keeping the other parameters equal to the default values defined in this section. 

\begin{figure}
    \centering
    \includegraphics[width=0.4\textwidth]{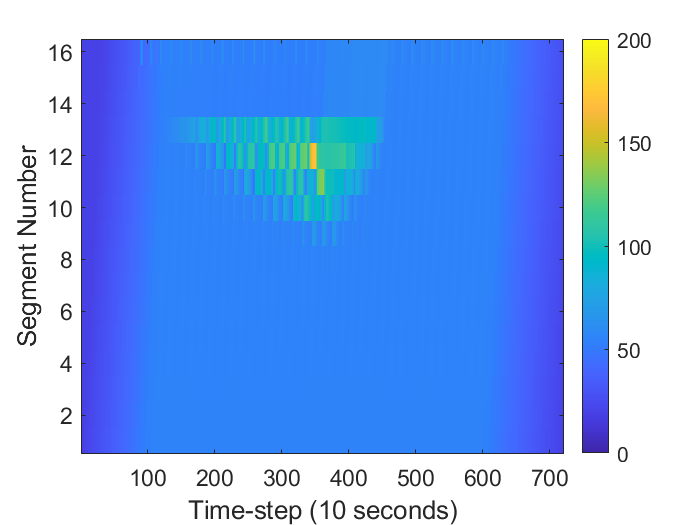}
    \includegraphics[width=0.4\textwidth]{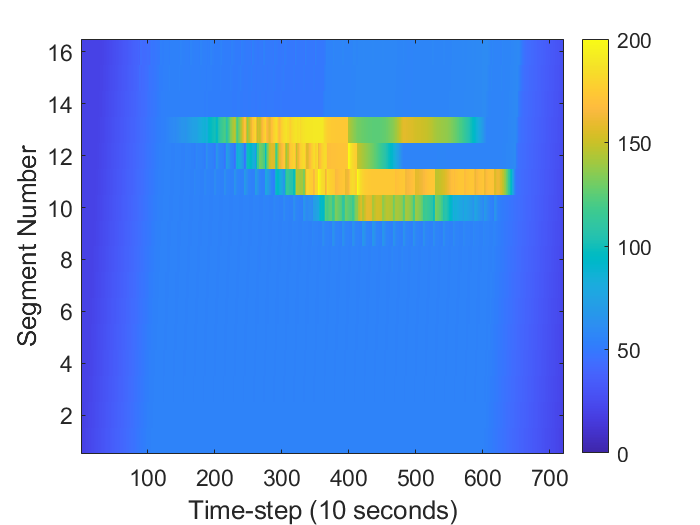}
    \caption{Density (veh/km) evolution on the highway stretch with [left] GN-LQR and [right] GN-LQRP controllers default parameters.}
    \label{f:LQR_default}
\end{figure}

For the $\textrm{GN-LQR}$ and $\textrm{GN-LQRP}$ algorithms, we select the weight matrices $\m Q$ and $\m R$ as $\begin{bmatrix}100\m{I_{N_L}} & \m 0 \\ \m 0 & \m 0_{N_{\mathrm{CAV}}}\end{bmatrix}$ and $\m{I_{N_{\mathrm{CAV}}}}$, respectively, where $N_L$ is fixed to the number of segments in the considered highway stretch while $N_{\mathrm{CAV}}$ varies with time depending on the number of CAV-platoons on the stretch at a given time. Here, the matrix $\m Q$ indicates that the weight is only applied to the density states and not the position of the CAV-platoons which are also states of the system. The penalty weight matrix $\m R'$ is set to $30 \m{I_{N_{\mathrm{CAV}}}}$. An equilibrium density of $59$ veh/km which is $1$ veh/km less than the critical density $\rho_c$ and an equilibrium speed of $99$ km/hr which is $1$ km/hr less than the free-flow speed $v_f$ is set for both controllers. In general, an equilibrium density of $\rho_c$ and a corresponding equilibrium speed of $v_f$ is desirable to achieve the maximum traffic flow. However, the equilibrium point is set slightly below these values to allow for numerical Jacobian calculation which requires the calculation of the nonlinear function at points around the equilibrium point. For speeds, $v_f$ is the upper bound, and for densities, the derivative is undefined at $\rho_c$ and changes sharply around that point thus making these exact values unusable for Jacobian calculation. In addition to the above settings, we set $N=3$ time steps. The maximum number of iterations for both GN-LQR and GN-LQRP is set to $1$ with an $\epsilon=0.001$. Figure \ref{f:LQR_default} shows the density evolution achieved by applying the GN-LQR and GN-LQRP controllers with the default set of parameters. The corresponding values of $\textrm{MS}$ are $94.5$ km/hr and $83$ km/hr, respectively. It can be observed from the figure that the default GN-LQR controller works well in reducing the congestion level on the highway stretch thus improving the $\textrm{MS}$ of the traffic. This is achieved by creating smaller controlled reductions in segment flows (by reducing the CAV-platoon speeds) upstream of the bottleneck segment (Segment $13$). Since the outflow of segments decreases with an increase in density above the critical density, reducing the flow of traffic in small amounts in the upstream segments thereby increasing their density in small amounts while preventing higher densities in the bottleneck segments results in overall  higher flows across all the segments. This is the underlying idea behind moving-bottleneck control which is correctly executed by the LQR-based controller. On the other hand, the GN-LQRP controller does not perform as well since the CAV-platoon speeds are not sufficiently reduced prior to reaching the bottleneck segment due to the penalty on speed changes. In the ensuing sections, we present a detailed analysis of the impact of various parameters of the LQR algorithm on its performance including cases in which GN-LQRP performs equivalently to the GN-LQR controller while also preventing large fluctuations in the control speed.

\subsubsection{Impact of horizon length $N$}\label{s:LQR_N}

\begin{figure}
    \centering
    \includegraphics[width=0.4\textwidth]{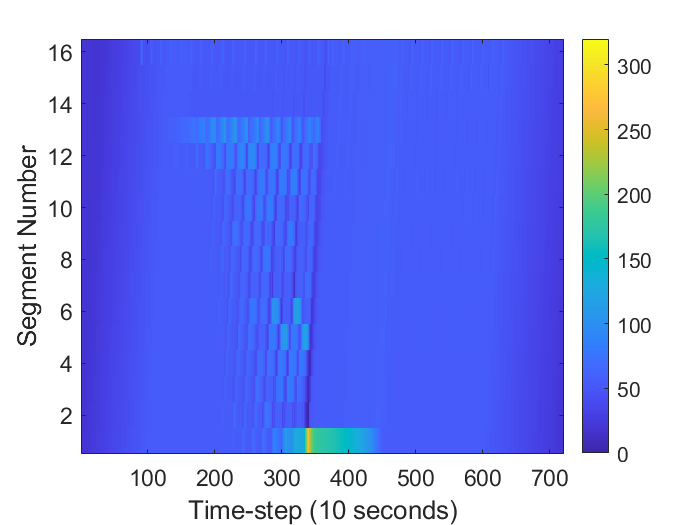}
    \includegraphics[width=0.4\textwidth]{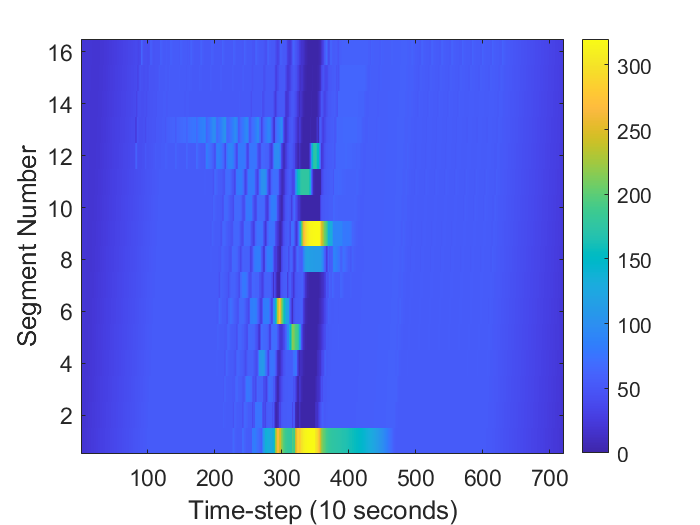}
    \caption{Density (veh/km) evolution on the highway stretch with GN-LQR controller with [left] $N=40$ time steps, and [right] $N=60$ time steps.}
    \label{f:LQR_40_60}
\end{figure}

\begin{figure}
    \centering
    \includegraphics[width=0.4\textwidth]{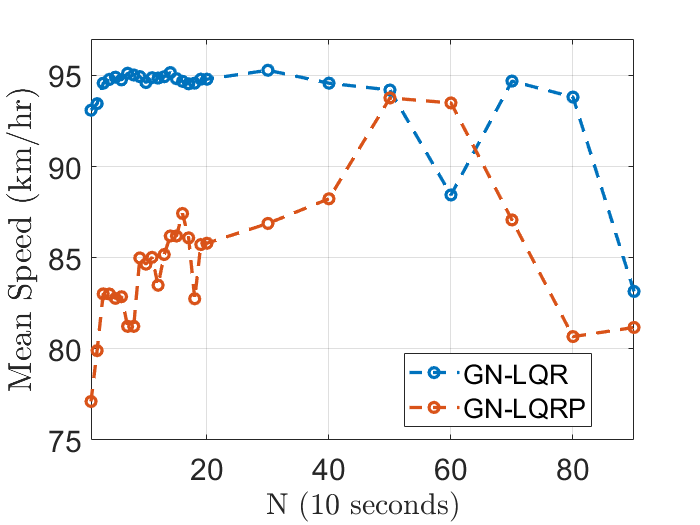}
    \caption{Variation in MS with horizon length N for GN-LQR and GN-LQRP.}
    \label{f:LQR_MS_vs_N}
\end{figure}

In this section, we test the impact of different values of the horizon length $N$ on the performance of the $\textrm{GN-LQR}$ and the $\textrm{GN-LQRP}$ controllers in terms of the achieved $\textrm{MS}$.
The value of $N$ is varied from $1$ to $90$ with increments of $1$ up to $20$ followed by increments of $10$. 
All other settings are as defined in Section \ref{s:lqr_default}. The corresponding values of $\textrm{MS}$ are presented in Figure \ref{f:LQR_MS_vs_N}. The GN-LQR algorithm performs well even with a small horizon length of $1$-time step improving the $\textrm{MS}$ by $15.6\%$ to $93\%$. The value improves further to $94.5\%$ by $N=3$ beyond which it does not improve much with $N$. It is observed that above $N=20$ the $\textrm{MS}$ also tends to decrease showing large dips at $N=60$ and $N=90$. 
On closely examining the plots of the density evolution, it is observed that above $N=20$, the controller prescribes the CAV-platoons to reduce their speeds at the upstream end of the highway stretch thus reducing the flow into the stretch eventually resulting in less vehicle accumulation at the bottleneck and improving the flow. The control makes sense since the controller is now able to look further into the future impact of each CAV-platoon and make the decision to reduce the platoon speeds sooner into the highway. See Figure \ref{f:LQR_40_60} [left] for the density evolution at $N=40$. However, this is not ideal, since spillbacks caused by congestion at the upstream end result in the creation of bottlenecks and the occurrence of capacity drop on previous links which is unaccounted for by this model. Additionally, at $N=60$ and $N=90$, there are further stoppages in the middle of the highway beside the stop at the entrance which results in reduced $\textrm{MS}$, for instance, see Figure \ref{f:LQR_40_60} [right] which shows the density evolution at $N=60$. This could be specific to the case and caused by a few stop decisions cascading into more stoppages. To avoid situations with jams created at the upstream end of the highway stretch, one solution is to avoid horizon lengths equal to or longer than the time taken by the CAVs to reach the bottleneck from the upstream end. Another solution is to attach additional weights to the input deviation term in the objective at the time of entry to ensure that CAVs enter the highway at free-flow speed and only reduce the speeds once sufficiently within the highway stretch to avoid spillbacks to previous links.

\begin{figure}
    \centering
    \includegraphics[width=0.4\textwidth]{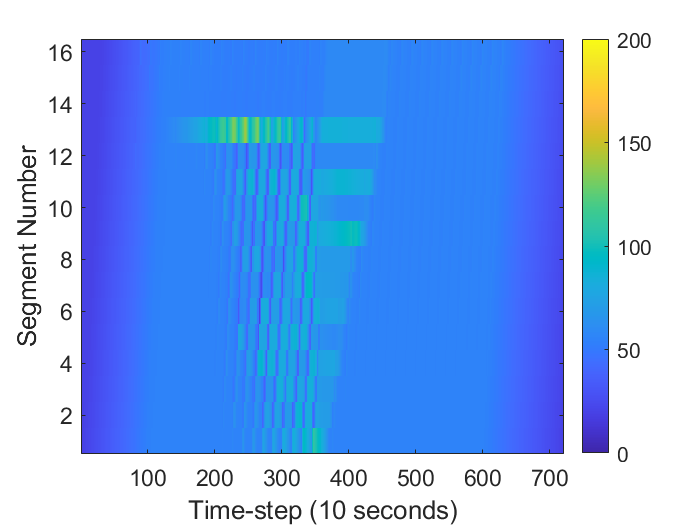}
    \caption{Density (veh/km) evolution on the highway stretch with GN-LQRP controller with $N=50$ time steps.}
    \label{f:LQRP_50}
\end{figure}

The performance of the GN-LQRP controller improves more substantially with increasing $N$ up to $N=60$ beyond which a degradation in its performance is observed. As also mentioned in the previous section, the reason for the poor performance of the GN-LQRP controller at low horizon lengths is the insufficient time for CAV-platoons to reduce their speeds due to the penalty on speed changes. As the horizon length increases, vehicles are able to start reducing their speeds sooner into the highway stretch and achieve enough speed reduction to reduce the impact of the bottleneck. Figure \ref{f:LQRP_50} presents the density evolution for the case with $N=50$ time steps showing the speed reduction which starts from the upstream end of the highway stretch. Note that this case does not result in spillbacks since Segment $1$ still has enough capacity to accommodate more vehicles from the previous link unlike the case in Figure \ref{f:LQR_40_60} [left] which reaches close to the maximum density in Segment $1$ for a brief period. Beyond $N=60$, we observe similar spillback situations with GN-LQRP as well as situations similar to Figure \ref{f:LQR_40_60} [right]. 

\begin{figure}
    \centering
    \includegraphics[width=0.4\textwidth]{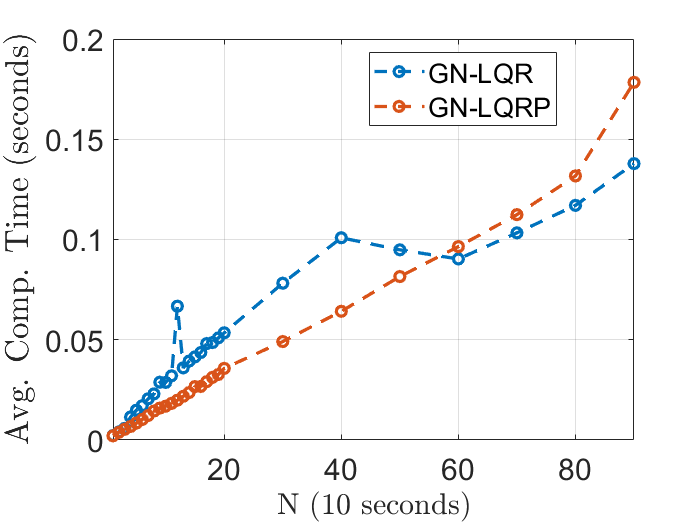}
    \caption{Variation in Average Computation Time (seconds) for each run of the controller with horizon length $N$ for $\textrm{GN-LQR}$.}
    \label{f:ACT_vs_N}
\end{figure}

The variation in ACT with an increase in the horizon length for both GN-LQR and GN-LQRP controllers is presented in Figure \ref{f:ACT_vs_N}. In the case of LQR-based controllers, the largest component of the computation time is dedicated to the computation of the derivatives of the nonlinear function. Therefore, as expected, the ACT increases almost linearly with the horizon length as the number of steps involving derivative calculation increase linearly. Deviation from the linear increase in ACT can be expected in some cases due to the accumulation of CAV-platoons on the highway which can increase the number of calculations per controller run or in cases where the CAV-platoons are blocked upstream of the highway reducing the number of calculations required. The order of magnitude for ACT is still a fraction of a second which makes it suitable for real-time control. 

\subsubsection{Impact of the number of iterations}
Here, we test the impact of changing the number of iterations of the GN-LQR algorithm on its control performance. The number of iterations is varied from $1$ to $10$ while keeping the other parameters equal to the default values. Table \ref{t:LQR_vs_iterations} presents the values of the $\textrm{MS}$ achieved at the different number of iterations. It is observed that the best performance of the GN-LQR controller is achieved at the number of iterations = $1$. From Table \ref{t:LQR_vs_iterations}, it can be seen that increasing the number of iterations degrades the performance of the controller. The main reason for this is a higher reduction in the speeds of CAVs with an increasing number of iterations. The iterations are intended to find a point where the controls used to obtain the derivative of the system are close to the controls obtained by using the derivative at which point the system is said to have converged. However, when the derivative is not significantly affected by the control inputs such that the direction of change in inputs does not change with the control inputs used to obtain the derivative, then the controls continue to grow/reduce over iterations and only converge at the lower/upper bound of the inputs. Since the optimal speeds at any iteration are used as the equilibrium speeds for the next iteration, then if in the first iteration, the optimal controlled speeds are below the equilibrium speed then in the following iterations, the control speeds will continue to be below the equilibrium speed for the respective iterations eventually resulting in a speed equal to the lower bound. In cases where the optimal speed in the first iteration is above the equilibrium speed, convergence is achieved in the first iteration itself as the controls are capped to the free-flow speed, and the initial equilibrium speed is already set close to the free-flow speed. Therefore, those cases result in the same control even with a higher number of maximum iterations. In general, as the number of iterations increases, the control speeds of the CAV-platoons tend to be lower than with maximum iterations = $1$ resulting in less than the best performance.

\begin{table}
    \centering
    \caption{Variation in MS with the number of iterations for GN-LQR.}
    \begin{tabular}{|c|c|c|c|c|c|c|c|c|c|c|}
        \hline
         $\#$iterations & 1 & 2 & 3 & 4 & 5 & 6 & 7 & 8 & 9 & 10\\
         \hline
         MS (km/hr) & 94.4101 & 90.4829 & 91.2102 & 88.5555 & 92.0806 & 88.3891 & 90.229 & 84.1028 & 83.3586 & 81.1909\\
        \hline
    \end{tabular}
    \label{t:LQR_vs_iterations}
\end{table}

\subsubsection{Impact of objective weights- $\textbf{Q}$, $\textbf{R}$, and $\textbf{R}'$}
\begin{figure}
    \centering
    \includegraphics[width=0.4\textwidth]{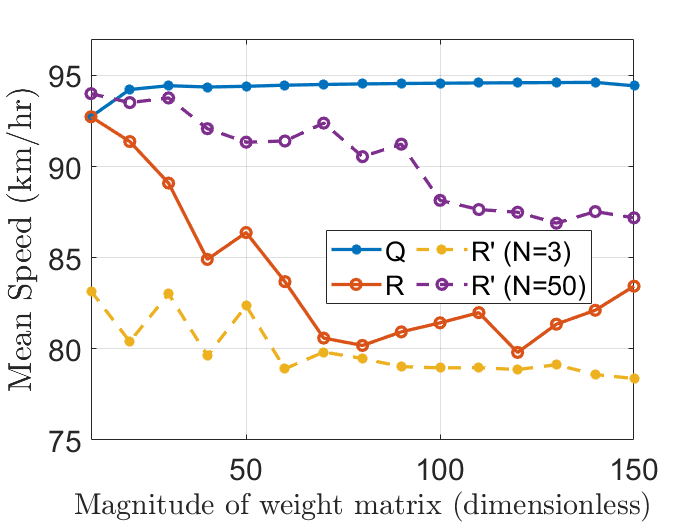}
    \caption{Variation in MS with LQR objective weight matrices for GN-LQR and GN-LQRP.}
    \label{f:LQR_MS_vs_Weights}
\end{figure}

\begin{figure}
    \centering
    \includegraphics[width=0.4\textwidth]{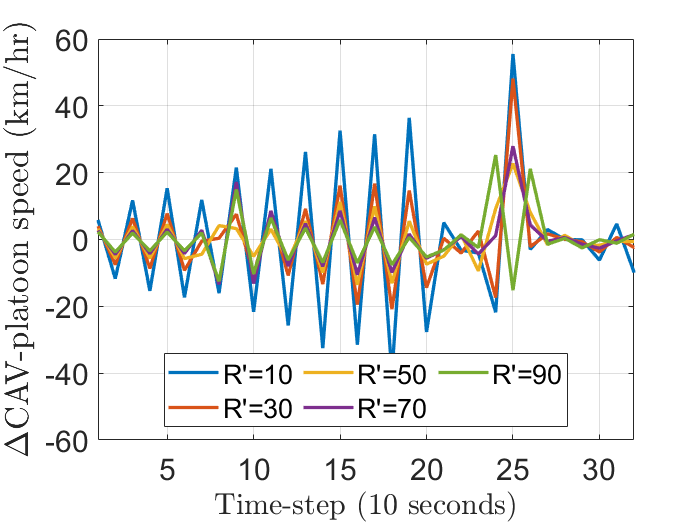}
    \caption{Change in control speed value over consecutive time steps at different values of penalty weight for GN-LQR with a penalty.}
    \label{f:CAV_speed_with_penalty}
\end{figure}

The objective of the LQR optimization problem defined in \eqref{e:LQR_objective} is different from the $\textrm{MS}$ metric used to judge the performance of the LQR-based controllers. Generally, the $\textrm{MS}$ is expected to improve if the controls cause the states to move close to the critical density. Larger weights on the error terms for either the states or the control inputs (or the change in control inputs in the case of GN-LQRP) incline the controller towards producing controls to reduce the corresponding errors. Here we assess the relationship between the various weights in the LQR objective function and the $\textrm{MS}$ obtained from the resulting control. For this analysis, the weights are defined in the form of diagonal matrices as $\m Q = \begin{bmatrix}w_Q\m{I_{N_L}} & \m 0 \\ \m 0 & \m 0_{N_{\mathrm{CAV}}}\end{bmatrix}$, $\m R = w_R\m I_{N_{\mathrm{CAV}}}$, and $\m R' = w_{R'}\m I_{N_{\mathrm{CAV}}}$ where $w_Q, w_R, w_{R'}\in\mathbb{R}$. The weights $w_Q, w_R,$ and $w_{R'}$ are individually varied from 10 to 150 with an increment of $10$ while keeping the other parameters equal to the default set of values. Figure \ref{f:LQR_MS_vs_Weights} presents the plots of $\textrm{MS}$ against the varying objective weights. It is observed that a larger difference between $\m Q$ and $\m R$ results in better MS. This is expected as the optimal control is to reduce the speed of the CAV-platoons before the bottleneck segment on the highway and a comparable weight on the speed (control input) error term prevents enough reduction in speed to improve the traffic flow. As also observed in Section \ref{s:LQR_N}, the performance of the GN-LQRP controller is worse than the GN-LQR controller for smaller values of horizon length due to insufficient reduction in speeds. Increasing the magnitude of $\m R'$, naturally results in further degradation in the performance as the reduction in speeds is further restricted. We also consider the variation in the values of $\m R'$ at $N=50$ time steps at which the performance is observed to be equivalent to the performance of the GN-LQR controller according to Figure \ref{f:LQR_MS_vs_N}. In this case, the performance is equivalent to the GN-LQR controller at the smaller values of $\m R'$ and decreases with an increase in $\m R'$ due to the same reason of insufficient reduction in speeds. Figure \ref{f:CAV_speed_with_penalty} presents a plot of changes in control speed over consecutive time steps for CAV-platoon $11$ which enters the highway stretch at $210$ time-step at different values of $\m R'$ with $N=50$ time steps.

\subsection{Comparison of GN-LQR and GN-LQRP with PI- and MPC-based controllers}
\label{s:comparison}
Two of the most common types of controllers implemented in traffic control systems are PI- and MPC-based controllers. PI-based controllers are much faster in computation and therefore adequate for real-time control, however, they do not offer the guarantee of optimal control. The MPC-based controller on the other hand guarantees optimality for a given horizon length, however when the system is nonlinear such as the system in the current study, the approach used to solve the problem is usually based on some meta-heuristic algorithm such as an evolutionary algorithm which can be inefficient due to repeating simulations and therefore unfit for real-time control. The LQR-based controllers investigated in the current study do not require performing the simulation several times to reach a solution, instead, it exploits the structure of the state-space matrices. In addition, the computed solution guarantees the optimality of the LQR objective which in this case ensures that the system states are as close to the critical density of the system as possible and therefore at the maximum flow.

The PI-based controller used for comparison in this study is implemented based on \cite{piacentini2019multiple}. The equation for obtaining the speed value at each time step is presented in \eqref{e:PI_speed_update}. Setting values for the controller gains $K_p$ and $K_I$ is a challenging problem in general. The authors in \cite{piacentini2019multiple} provide certain fixed gain values for the PI-based controller but do not present a formal method to derive them. Since the setup is slightly different from the one used in \cite{piacentini2019multiple}, we obtain optimal gains for the scenario in this study by setting up a nonlinear optimization problem with the objective of minimizing the MS. The details of the fitting are presented in Appendix \ref{a:PI}. The study \cite{piacentini2019multiple} also prescribes a lower bound of $60$ km/hr for the control speed. This lower bound is implemented by projecting the controller speed to within the bounds. A reason for this lower bound is to avoid the sudden large drops in speed to extremely low values of speed which may lead to accidents due to the low reaction time of drivers. However, with increased connectivity and autonomy in vehicles, it is possible to expect no plausible limit to what the speeds can be dropped to. In this study, we, therefore, also test the controller with and without a lower bound on the speed.

The MPC-based controller is also implemented according to \cite{piacentini2019highway}. The objective and constraints are set exactly as in \cite{piacentini2019highway} and are presented in Appendix \ref{a:MPC} for reference. The horizon length for the MPC-based controller is set to $20$ time steps as in \cite{piacentini2019highway}. Again, a lower bound for the value of control speeds equal to 60 km/hr is prescribed. This lower bound can be naturally incorporated into the MPC-based controller as a constraint. 

The LQR-based controller investigated in this study does not inherently allow for a lower bound similar to the above controllers. Instead of applying a lower bound similar to the PI-based controller, in this work, we use the GN-LQRP controller which applies a penalty on changes in the speeds. This is different from the implementation for PI-based and MPC-based controllers which do not account for changes in consecutive time steps. All the parameters for the LQR-based algorithms are set to their default values as defined in Section \ref{s:lqr_default} with the exception of the horizon length for GN-LQRP which is set to $50$ time-steps as it is observed to be the best setting for GN-LQRP in Section \ref{s:LQR_N}.

\begin{figure}
    \centering
    \includegraphics[width=0.4\textwidth]{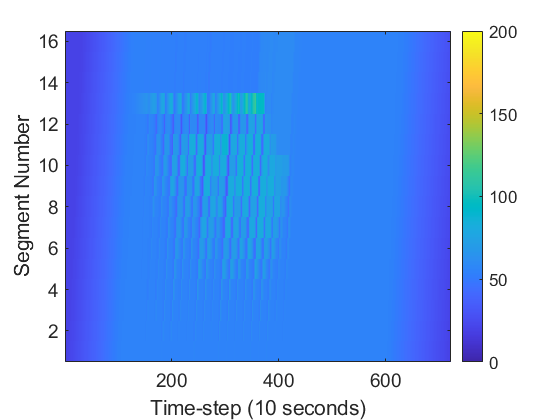}
    \includegraphics[width=0.4\textwidth]{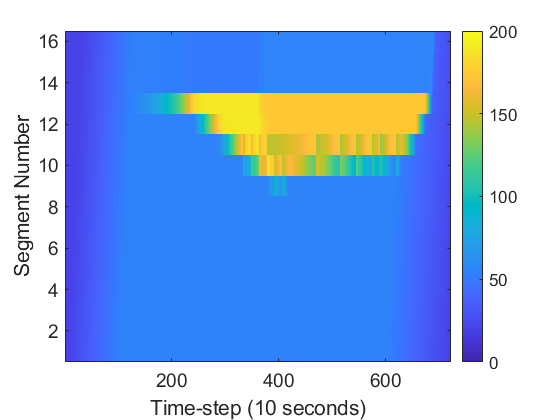}
    \includegraphics[width=0.4\textwidth]{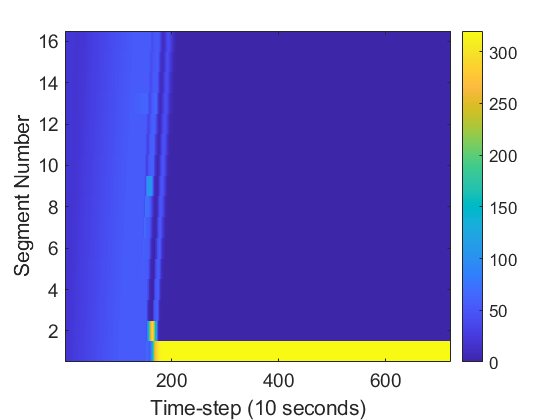}
    \caption{Density (veh/km) evolution on the highway stretch with PI-based control with [top left] optimal gains and lower bound of $60$ km/hr, [top right] optimal gains and no lower bound, and [bottom] arbitrary gains and no lower bound.}
    \label{f:PI_controller_density}
\end{figure}

The obtained values for $K_P$ and $K_I$ for the PI-based controller with a lower bound of 60 km/hr on the controlled speeds are $0.7944$ and $0.1091$, respectively. The values obtained without a lower bound are $0.7908$ and $-8.9832$, respectively. Figure \ref{f:PI_controller_density} presents the density evolution plots for the PI-based controllers with and without a lower bound. The TTT, TTD, and MS for the tested controllers are presented in Table \ref{t:compare_controls}. It is observed that the PI-based controller with a lower bound reduces the impact of the bottleneck by slowing down the traffic approaching the bottleneck thus creating smaller jams upstream of the bottleneck and reducing the density of the cells at the bottleneck which reduces the effect of capacity drop and improves the \textrm{MS}. On the other hand, the optimal controller gains without a lower bound result in no improvement in the MS over the uncontrolled case. To further understand the ineffectiveness of the PI-based without a bound, we set the controller gains to the values in \cite{piacentini2019multiple} instead of their optimal values. Figure \ref{f:PI_controller_density} also presents the plot of density evolution for this new setting of the PI-based controller. It is seen that the obtained control, in this case, is to stop all the vehicles at the entrance which is possible in this case since the speeds can drop to $0$ km/hr. This makes sense for reducing the error term of the controller which only penalizes segments ahead of the controlled platoons that have a density above the critical density and in this case, the densities for all the segments ahead of the first segment are zeros therefore there is theoretically no error for the controller. However, this is not ideal for the traffic flow since it only reduces congestion on the current highway stretch at the expense of creating a spillback upstream of the stretch that results in congestion upstream. It results in a small MS value of $38.7 \%$ despite a small TTT value of $413.1$ since the TTD value becomes very small equal to $16,002$ due to a smaller number of vehicles entering the stretch. Therefore the restricted speeds offer better results in this case in terms of our evaluation metrics.

\begin{table}
    \centering
    \caption{Comparison metrics over different traffic control scenarios for the given example. Computation time (CT in seconds) refers to ACT for MPC- and LQR-based controllers and offline computation time for PI-based controllers.}
    \begin{tabular}{|c|c|c|c|c|}
        \hline
         Scenario & TTT & TTD & MS & CT \\
         \hline 
         No Control & 1,019 & 78,998 & 77.5 & -\\
         \hline
         PI (lower bound 60 km/hr) & 820.8 & 78,741 & \textbf{95.9} & 7.1051\\
         \hline
         PI (no lower bound) & 1017.5 & 78,741 & 77.3 & 1.5982\\
         \hline
         MPC (lower bound 60 km/hr) & 837.8 & 78,727 & 93.6 & 2.8729\\
         \hline
         MPC (no lower bound) & 817.1 & 78,727 & 92.7 & 3.3722\\
         \hline
         GN-LQR ($N=3$) & 832.5 & 78,741 & 94.5 & \textbf{0.0058}\\
         \hline
         GN-LQRP ($R'=30 \m I, N=50$) & 839.7 & 78,741 & 93.7 & 0.0884\\
         \hline
    \end{tabular}
    
    \label{t:compare_controls}
\end{table}

\begin{figure}
    \centering
    \includegraphics[width=0.4\textwidth]{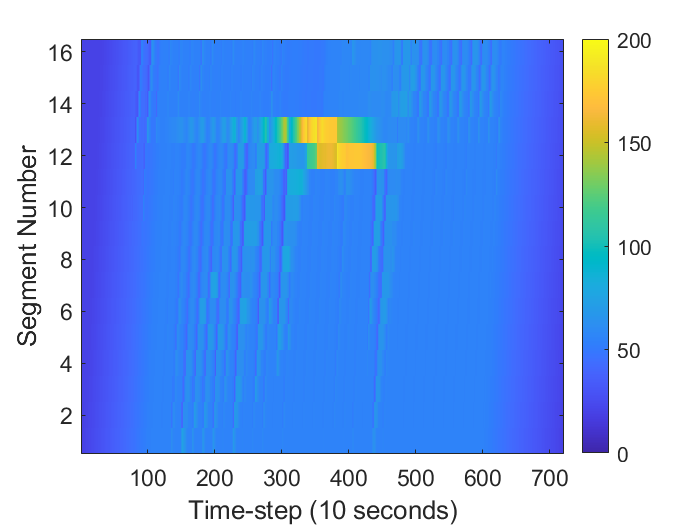}
    \includegraphics[width=0.4\textwidth]{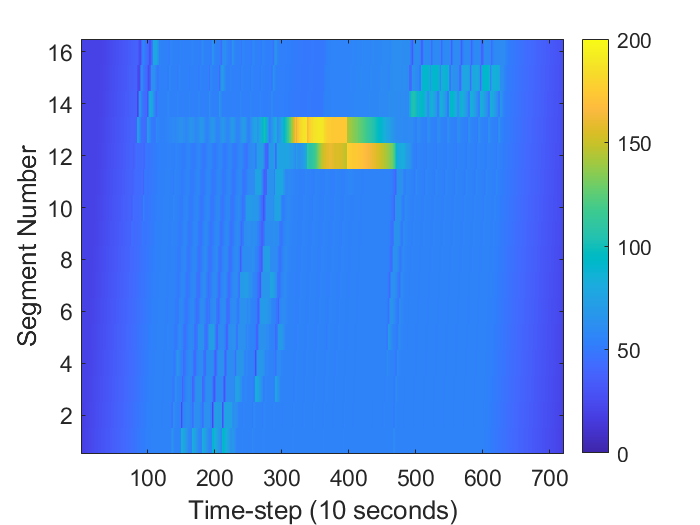}
    \caption{Density (veh/km) evolution on the highway stretch with MPC-based control with a [left] lower bound of 60 km/hr, and [right] no lower bound.}
    \label{f:MPC_controller_density}
\end{figure}

Figure \ref{f:MPC_controller_density} presents the density evolution plots for the MPC-based controllers with and without a lower bound which are similar in this case. The MPC-based controller tries to minimize the $\textrm{TTT}$ while maximizing the outflow from the bottleneck and keeping the density of the bottleneck segment close to the critical density. Therefore, in both cases, it tries to reduce the density at the bottleneck to reduce the impact of the capacity drop rather than stopping all vehicles upstream of the stretch as in the case of the PI-based controller. In the case of MPC, while stopping all vehicles at the entrance would still minimize the \textrm{TTT}, it potentially creates a larger gap between the critical density and the density of the bottleneck segment and reduces the outflow from the bottleneck thereby making such a solution sub-optimal. 

\begin{figure}
    \centering
    \includegraphics[width=0.4\textwidth]{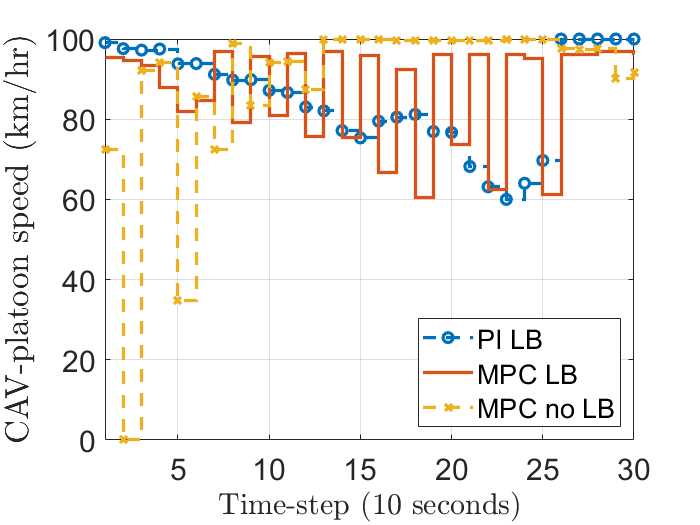}
    \includegraphics[width=0.4\textwidth]{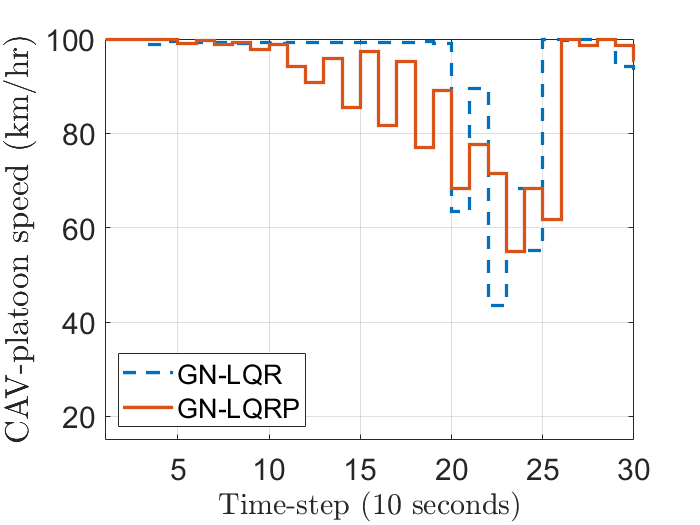}
    \caption{CAV platoon speed profile for platoon 11 with [left] PI-based controller with lower bound, and MPC-based controller with and without lower bound, and [right] GN-LQR controller and GN-LQRP controller with $\m R' = 30\m I$.}
    \label{f:controller_speeds}
\end{figure}

The density evolution plots for the LQR-based controllers are presented in Figures \ref{f:LQR_default} and \ref{f:LQRP_50} and the evaluation metrics are reported in Table \ref{t:compare_controls}. It is observed that the performance of the LQR-based controllers, in this case, is comparable to both implementations of the MPC-based controller and the lower bounded implementation of the PI-based controller, while it clearly outperforms the PI-based implementation without a lower bound. Figure \ref{f:controller_speeds} presents the speed profile of CAV-platoon $11$ that enters the highway at time step $210$ for all the controllers. Note that the speed profiles for the PI-based and the GN-LQRP controllers are the smoothest with a gradual reduction in speeds. The original GN-LQR controller and the MPC-based controller without a lower bound result in large abrupt changes in speeds for the platoon within a time step. While theoretically, such variations in speed are possible within a time step (which is equal to $10$ seconds) they can be unsafe under high reaction times. While the MPC-based controller with a lower bound prevents as high of speed fluctuations as in the GN-LQR and MPC-based without a bound, we can still observe fluctuations close to $40$ km/hr which is the shift from the maximum speed to the lower bound value. This shows that the lower fluctuations are not a property of the controller (as in the case of GN-LQRP) but an artifact of the lower bound which is arbitrarily imposed.

Table \ref{t:compare_controls} also presents the computation time ($\textrm{CT}$) in seconds for all the controllers which refers to the ACT in the case of MPC- and LQR-based controllers and to the offline computation time for gain calculation in the case of PI-based controllers. The PI-based controller is the fastest of the controllers as the gain computations are performed offline and there is virtually no computation time for the controller in real-time. Although in cases when the offline gains do not work as expected due to different realization of the traffic conditions than expected, then real-time computation of gains may be required similar to the MPC-based controller with a finite horizon for which the traffic conditions can be reliably known. Since the underlying problem is nonlinear, it will require solving a nonlinear optimization problem in real-time which would also be computationally expensive, only less expensive than MPC-based due to the lower number of control variables which in this case would just be the values of the two gains. The computation time for the LQR-based controllers mostly comprises the time to compute the derivatives of the state space equation with respect to the equilibrium states and inputs. As seen in Figure \ref{f:ACT_vs_N}, the computation time of the GN-LQR almost linearly increases with $N$ as the derivative calculations increase. Although since gradients can be computed with very few computations of the nonlinear function of the state-space model, this time requirement is negligible and therefore the overall time for obtaining controls from the LQR-based controller is also quite small. As expected, the computation time for the MPC-based controller is the highest of all the controllers due to the nonlinear optimization performed every time the controller is run. Note that computation times can vary significantly based on implementation and the authors do not claim their implementation of the controllers to be the most efficient. However, given that LQR-based and PI-based computations are expected to be faster due to the presence of only algebraic computations as compared to MPC-based which requires performing the simulation several times to solve the nonlinear optimization problem, the results for the computation times do serve to validate the hypothesis about the expected computational differences between MPC-based and the other controllers.

\subsection{Microsimulation-based case study}
\label{s:microsimulation}
In this section, we reproduce the traffic control scenarios presented in Section \ref{s:scenario_and_metrics} using a realistic microscopic traffic simulator and use it to test the performance of the proposed GN-LQR and GN-LQRP control algorithms under a realistic setting. The existing PI- and MPC-based controllers are also tested in microsimulation for the same setting. The micro-simulation is performed using TransModeler $6.1$ \cite{yang1996microscopic,balakrishna2009large} while the control algorithm is implemented using MATLAB R$2021$b. The GISDK \cite{tsm} API in TransModeler is used to interact with the controller. All processes related to the microsimulation-based analysis are run on a $64$-bit Windows $10$ with a $2.3$GHz IntelR CoreTM i$7$-$11800$H CPU and $16$GB of RAM.

\subsubsection{Simulation and control pipeline}

\begin{figure}
    \centering
    \includegraphics[width = \textwidth]{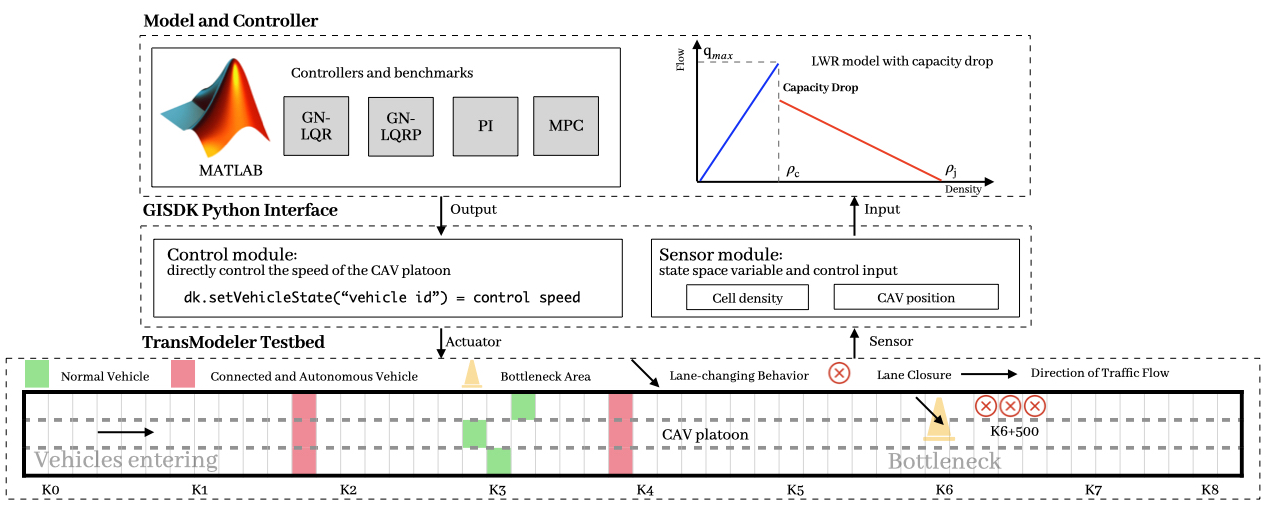}
    \caption{The simulation framework developed in TransModeler to validate the controller.}
    \label{f:cosim}
\end{figure}

Note that in the microsimulation, the proposed controllers are tested in a scenario involving a multi-lane highway by controlling CAV-platoons formed by CAVs positioned side-by-side, acting as a rolling roadblock.
The simulation framework is shown in Figure \ref{f:cosim}, consisting of four important modules: the TransModeler testbed, GISDK Python interface, CTM-based state-space model and controller, and visualization. The framework closes the loop for the state-feedback control, where TransModeler provides the testbed to simulate realistic traffic conditions. The simulation framework supports the trajectory-level traffic analysis, which helps us better understand and explain the control mechanisms of the proposed controllers. The ensuing sections provide a detailed overview of the simulation settings.

\subsubsection{Network and demand}
\begin{figure}
    \centering
    \includegraphics[width=0.32\textwidth]{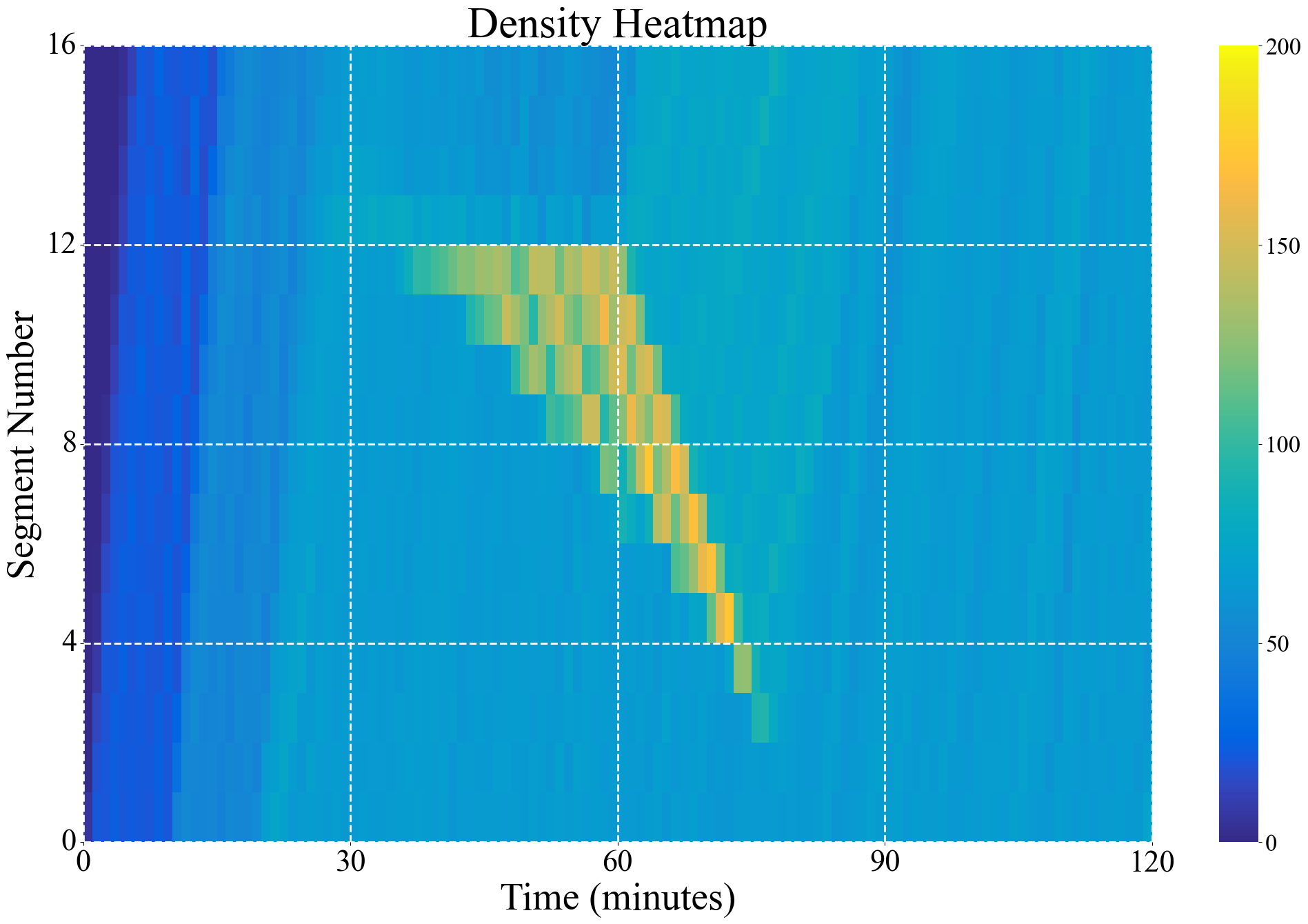}
    \includegraphics[width=0.32\textwidth]{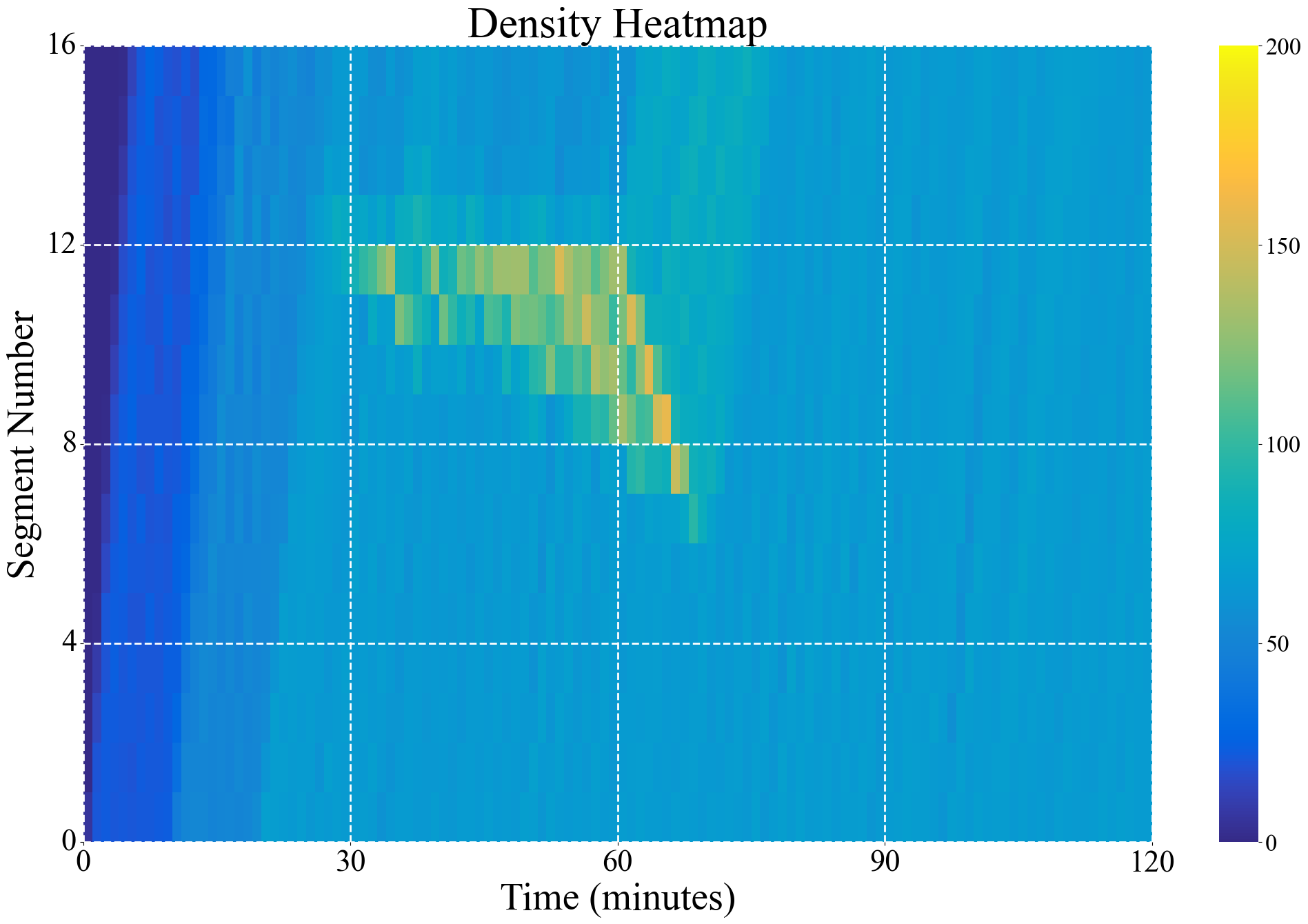}
    \includegraphics[width=0.32\textwidth]{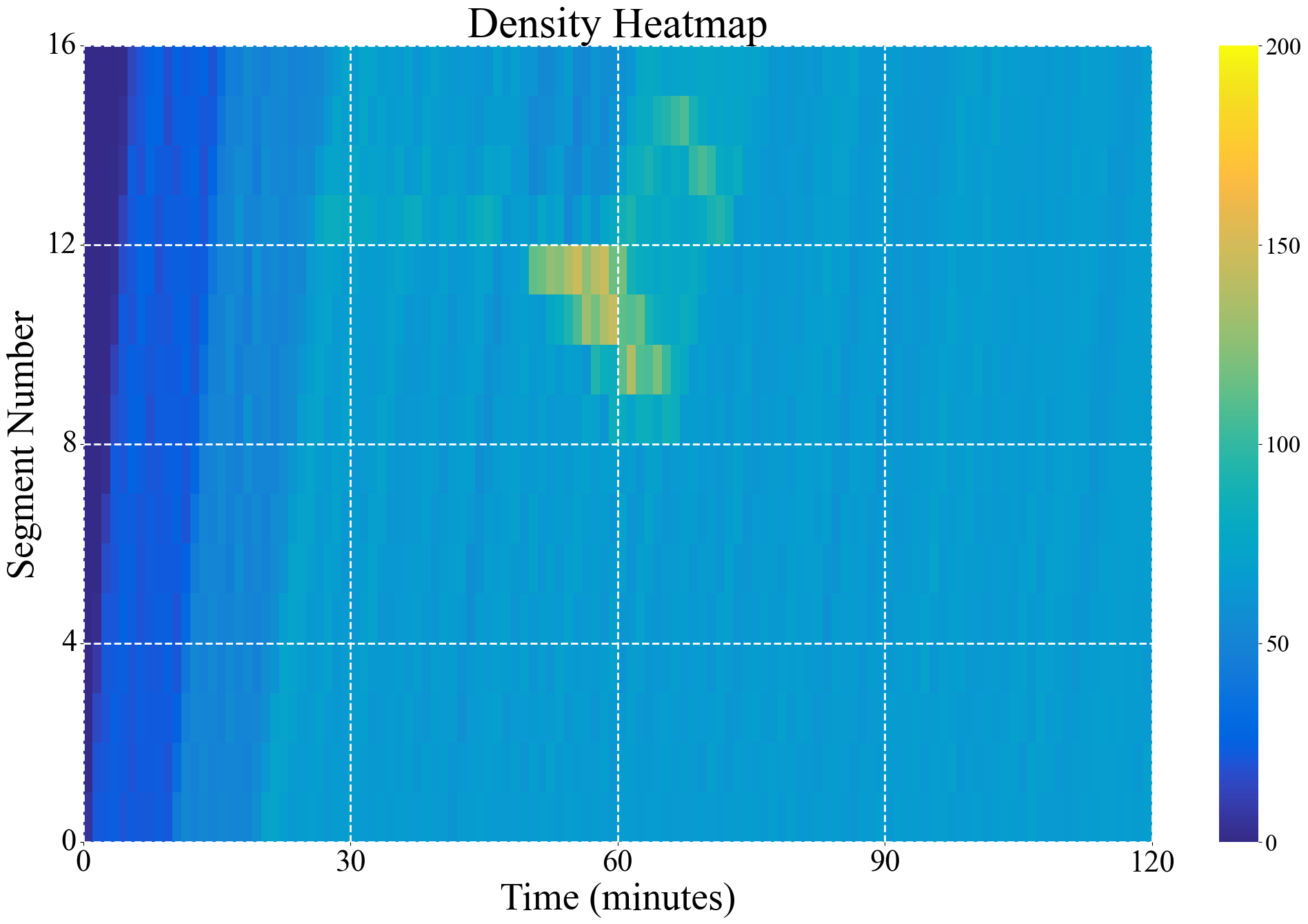}
    
    \includegraphics[width=0.33\textwidth]{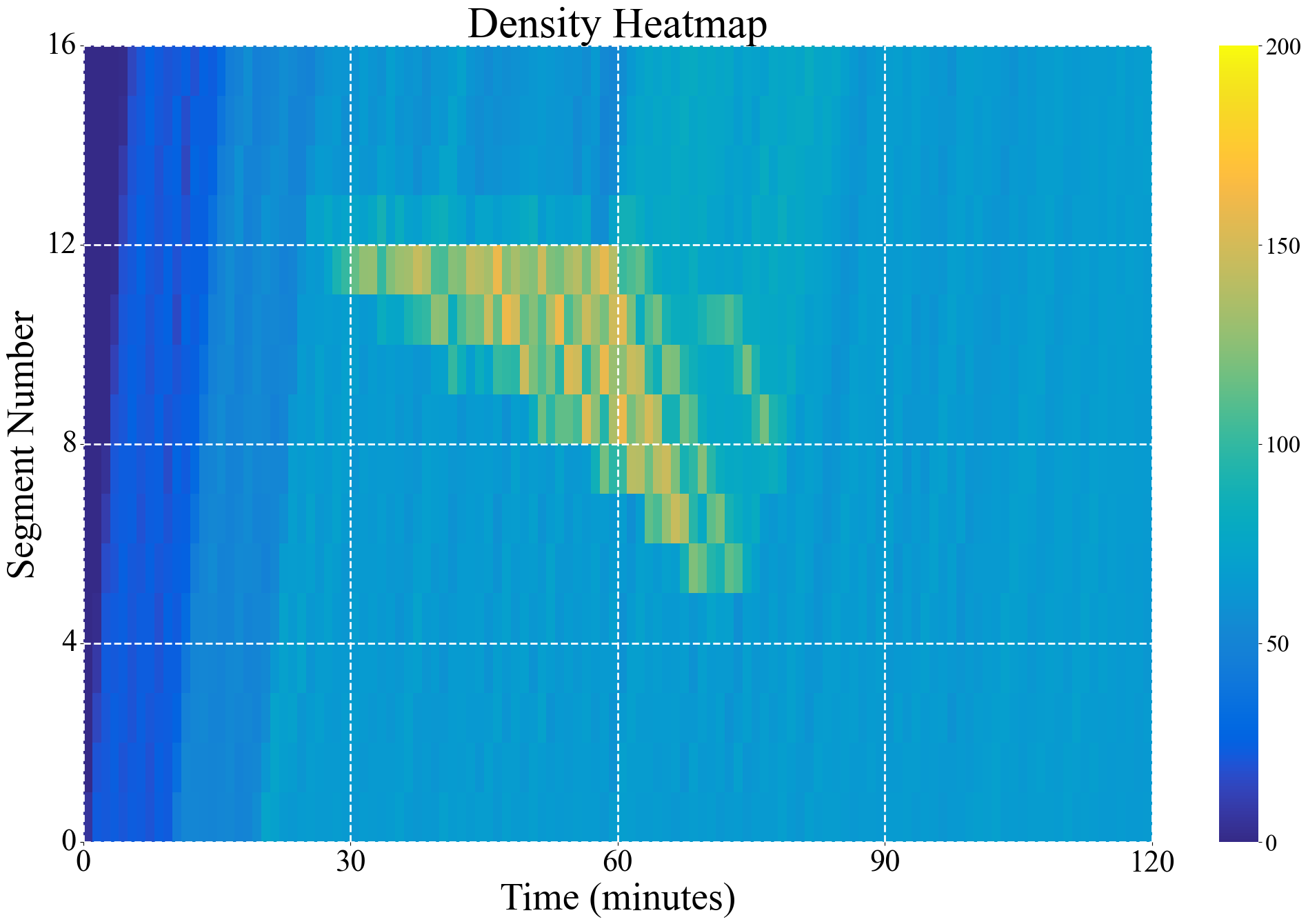}
    \includegraphics[width=0.33\textwidth]{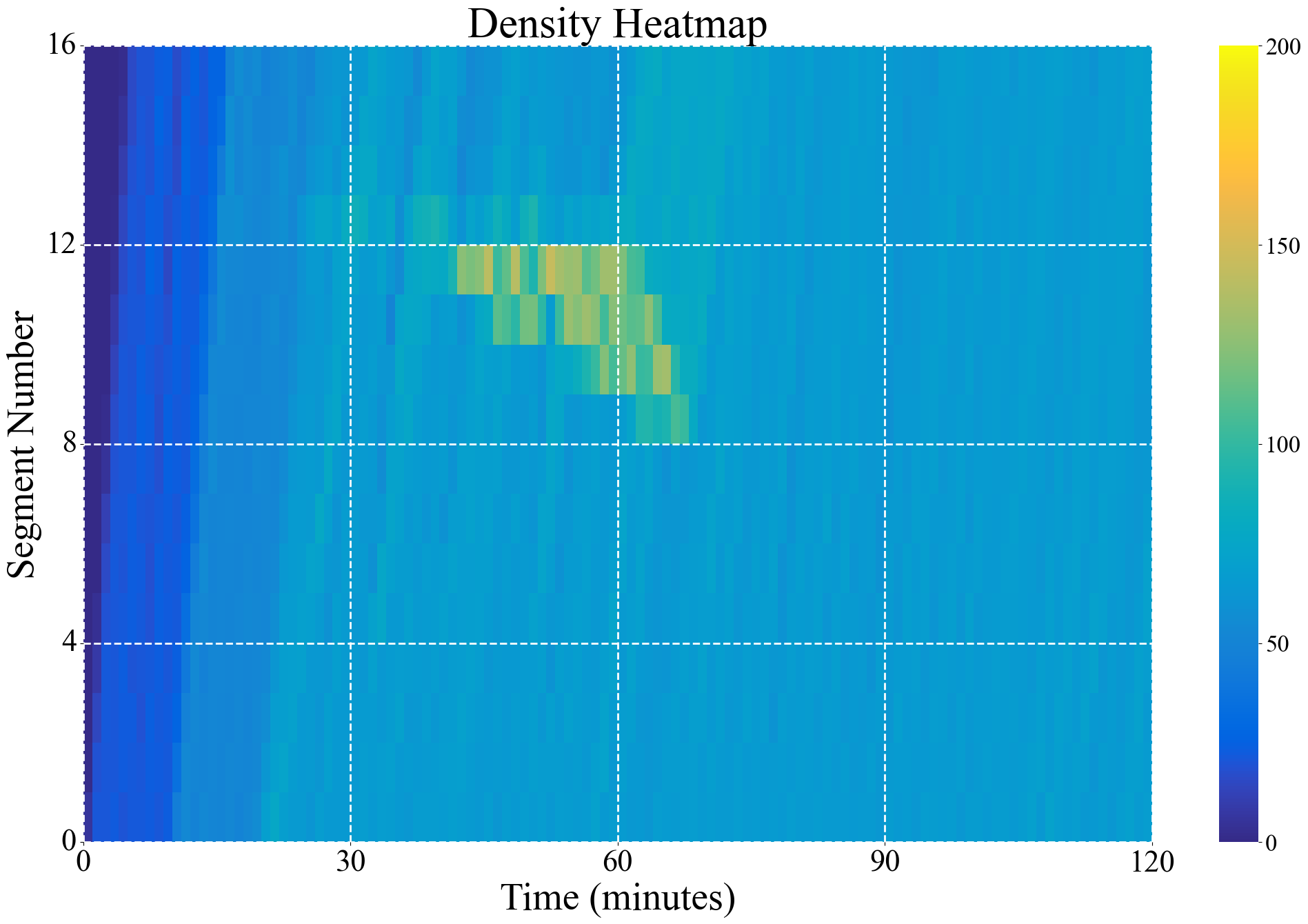}
    \caption{Density (veh/km) evolution on the highway stretch in microsimulation for the different scenarios: [top left] no control with bottleneck, [top middle] with LQR control, [top right] with LQRP control, [bottom left] with PI control, [bottom right] with MPC control.} 
    \label{f:ms_density}
\end{figure}

The tested road network is composed of an $8$ km long highway stretch with $3$ lanes. The same is depicted in Figure \ref{f:cosim}.  In  Figure \ref{f:cosim}, the origin of the considered road stretch is marked using meter markers as K0, and the end of the stretch is marked as K8. To capture the real demand and supply conditions  and effectively form a CAV platoon before they enter the considered stretch of the roadway, two buffer zones of length $1$ km each are established at both the beginning and the end of the highway segment, hence a $10$ km long highway is simulated.
The capacity of the highway stretch is set at $2000$ veh/lane-hr. 
The speed limit for the entire stretch is $100$ km/hr.
The total simulation duration is $2$ hours. 
The demand profile is the same as in Figure \ref{f:demand_profile}.
The microscopic simulation parameters are carefully tuned to reproduce a bottleneck and capacity drop which are consistent with the scenario described in Section \ref{s:scenario_and_metrics}. A bottleneck is simulated on Segment $13$ with the help of a lane-changing guide signal with a $30$\% compliance rate. This lane-changing guidance system is implemented in the inner lane (top lane in Figure \ref{f:cosim}) and is expected to prompt $30$\% of the traffic to switch lanes, resulting in an observed $10$\% decrease in outflow for Segment $13$ which is the same as implemented in the macrosimulation-based case study. Interested readers are referred to \cite{ahmed1999modeling} for more details on the dynamics model, parameter settings, and tuning for TransModeler, and move to the Appendix \ref{a:car_following} to see more details on the models.
 In this case, the bottleneck begins at the start of the simulation and ends at the end of the first hour. Figure \ref{f:ms_density} [top left] shows the evolution of density with a bottleneck without control. 

\subsubsection{Control actuator}
The control actuators in this control system consist of vehicles (CAV-platoons) on the highway. Each CAV-platoon comprises of three vehicles moving side-by-side on the three lanes and acting as a rolling roadblock thus blocking all traffic behind them and not letting any vehicles overtake them. Since the road is blocked by the CAV-platoon, the control speed of the platoon will be mandatorily enforced on the upstream traffic, which is the key to the effectiveness of the controllers. These CAV-platoons are dispatched to the highway stretch using the $\texttt{AddVehicle()}$ function in TransModeler. The origin, destination, lane, and speed of the platoon  vehicles are then customized based on the simulation settings. The three CAVs are positioned separately on Lane $1$, Lane $2$, and Lane $3$ (where Lane $1$ is the inner lane and Lane $3$ is the outer lane) at the same location. The IDs of the CAV-platoons are recorded and monitored for sensing and control purposes. Upon activation of the controller, the speed of the CAV-platoon is directly regulated to the recommended speed using the $\texttt{SetVehicleInfo()}$ function in TransModeler. The state of the CAV-platoon is tracked using the $\texttt{GetVehicleInfo()}$ function in TransModeler, and the trajectory of the platoon is monitored to serve as both the control input and the input for result visualization.
\subsubsection{Sensors deployment and output}
 In this simulation, the sensors are categorized into two types: fixed sensors and mobile sensors. The fixed sensors are positioned along the highway diagram, which is divided into $16$ segments of equal length, each spanning $0.5$ km same as the space discretization in the state-space model of the controller. The output of the sensors includes the density of each segment. The simulation assumes that the density of each segment can be directly measured, instead of having to be estimated. The mobile sensors refer to the CAV-platoons that are dispatched to the road. The position of the CAV-platoons is used as the sensing input for the controllers.
\subsubsection{Metrics calculation} For the microsimulation analysis, metrics including TTT, TTD, and MS are calculated using vehicle trajectories, such that $TTT=\sum _{i=1}^{N_V} t_i$, and $TTD=\sum _{i=1}^{N_V} x_i$, where $x_i$ and $t_i$ are the travel distance and travel time for vehicle $i$, and $N_V$ is the total number of vehicles loaded in the microsimulation. To generate the density dynamics heatmap, according to Edie's definition, the traffic density ($\rho_{\text{Edie}}$), flow ($Q_{\text{Edie}}$), and speed ($V_{\text{Edie}}$) \cite{edie1961car,edie1963discussion} can be defined by the following equations:
\begin{equation}
    \rho_{\text {Edie }}=T^{\mathrm{tot}} / A, \quad Q_{\text {Edie }}=X^{\text {tot }} / A, \quad V_{\text {Edie }}=Q_{\text {Edie }} / \rho_{\text {Edie}},
    \label{eq:fd}
\end{equation}
when $t-\frac{\Delta t}{2}\leq t_i \leq t+\frac{\Delta t}{2}, ~x-\frac{\Delta x}{2}\leq x_i \leq x+\frac{\Delta x}{2}$. The 4 parameters $t, ~\Delta t, ~x, ~\Delta x$ bound a spatio-temporal box that contains multiple trajectory points, where $A = \Delta x \times \Delta t$. Here, $T^{tot}$ is the total travel time of the vehicles in the bounded boxes, and $X^{tot}$ is the total travel distance. Finally, MS is calculated in the same way as described in Section \ref{s:scenario_and_metrics}.

\subsubsection{Process update frequencies} Three types of update frequencies are considered in this work which are described as follows:
(1) \textbf{Simulation step update frequency} (10 Hz) is the rate at which the simulation progresses. In other words, it's the number of times the simulation updates per second. At a frequency of 10 Hz, the microsimulation updates 1 time per 0.1 seconds. Each update would correspond to a "step" in the simulation, during which the state of the simulation could change based on the inputs and the underlying model.
(2) \textbf{Controller update frequency} (0.1 Hz) is the rate at which the state of the controlled system (which could include various parameters or variables representing the highway dynamics and CAV platoon) and the control action (in this paper, is the speed of the CAV platoons) is updated. With a frequency of 0.1 Hz, the state and the control action are updated every 10 seconds.
(3) \textbf{Controlled CAV-platoon speed update frequency} (1 Hz) is the rate at which the controlled speed (the output from the controller) is actuated by the CAV platoon. At a frequency of 1 Hz, the suggested speed from the controller is actuated by the CAV-platoons every 1 second. Before the suggested speed is updated, the CAV-platoon will forward at the same speed. This ensures that the platoon is receiving relatively frequent updates about the speed it should be traveling at.

\begin{table}[h]
\centering
\caption{Evaluation metrics for different scenarios tested in the microsimulation.}
\begin{tabular}{|c|c|c|c|}
\hline
Scenarios & TTT   & TTD    & MS  \\ \hline
No bottleneck &   949 & 78,708 & 82.94 \\ \hline
No control with bottleneck & 1,089 & 79,223 & 72.75  \\ \hline
GN-LQR ($N=3$)  & 996      &  79,111      &   79.43\\ \hline
GN-LQRP ($N=50$)   & 1,016      &   76,903  &  75.67   \\ \hline
GN-LQRP ($N=30$)   & 1,021      &   79,368  &  77.73   \\ \hline
GN-LQRP ($N=10$)   & 974     &   79,286  &  \textbf{81.43}   \\ \hline
PI (lower bound 60 km/hr)           & 1,026      &   79,191     &   77.16    \\ \hline
MPC (lower bound 60 km/hr)         &   983   & 79,206      & 80.58   \\ \hline
\end{tabular}
\label{t:micro_comp}
\end{table}

\begin{figure}
  \centering
  \includegraphics[width=1\textwidth]{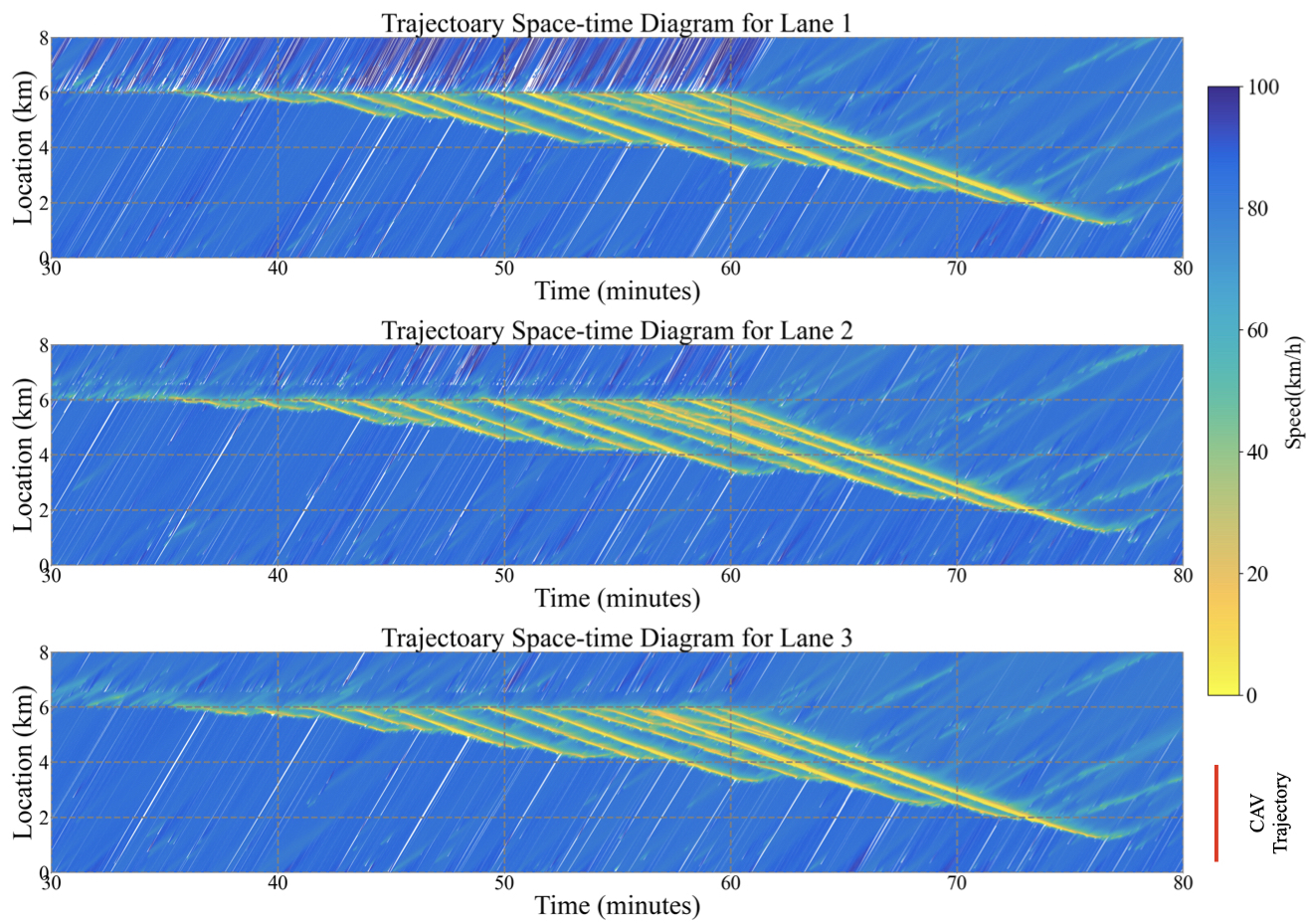}
  \caption{Vehicle trajectory space-time diagram for the uncontrolled scenario (the color of each trajectory point reflects the speed of the vehicle at the time): [top] the left lane, [middle] the middle lane, [bottom] the right lane.}
\label{f:micro_uncontrolled_traj}
\end{figure}

\subsubsection{Results and discussion}
\label{s:results_discussion}
In this section, we discuss the results obtained from the implementation of the aforementioned controllers namely GN-LQR, GN-LQRP, PI- and MPC-based. The parameter settings used for the various controllers are the same as in Section \ref{s:comparison}. For the PI- and MPC-based controllers, we only test the case with a lower bound of $60$ km/hr on the control speed. For GN-LQRP, we also test two additional runs with different values of $N$ as described later in this section. 
Table \ref{t:micro_comp} presents the values of the evaluation metrics obtained from the microscopic traffic simulation both in the presence and absence of a bottleneck on the highway and in the presence of control using the aforementioned algorithms to mitigate the impact of the bottleneck. The computation time for the controllers is omitted in this table as the controllers' implementation is the same as in the above sections and there is no significant difference in computation time. From the table, it can be observed that for the case without a bottleneck, unlike the macroscopic scenario described in Section \ref{s:scenario_and_metrics}, the MS is 82.94 rather than 99.9 (which is almost equal to the free-flow speed). This is because in microsimulation, even though the desired speed of traffic is the free-flow speed which is the same between the macrosimulation and microsimulation, the modeled behavior of drivers and vehicle-vehicle interactions can result in reduced speeds on various occasions in the simulation. The MS is further reduced to 72.75 in the presence of the bottleneck as the bottleneck causes an increase in the TTT due to additional lane changing resulting in a slowing down of upstream traffic. The slight increase in the TTD in the case with no bottleneck is because of the additional CAV-platoons ($36$ platoons, that is 108 vehicles) that are loaded onto the network in this case but are left uncontrolled. Figure \ref{f:ms_density} presents the evolution of density in the controlled scenarios for the different controllers similar to those presented for the macroscopic simulations. 
Plots of the trajectories of vehicles in the simulation from $30$ to $80$ minutes for the three lanes are presented in Figure \ref{f:micro_uncontrolled_traj}. The remaining simulation duration is omitted from the plot as it mostly contains traffic in a free-flowing state. Notice, that in the uncontrolled case, the slowing down of vehicles at the bottleneck extends upstream up to around the $2$ km mark. The increase in the length of the jam results from the slowdown of approaching vehicles behind the already slowed-down vehicles from lane changing at the bottleneck. As the resulting slowed-down vehicles eventually arrive at the bottleneck, they are again required to change lanes which causes the jam to continue in time.
  
\begin{figure}
      \centering
      \includegraphics[width=1\textwidth]{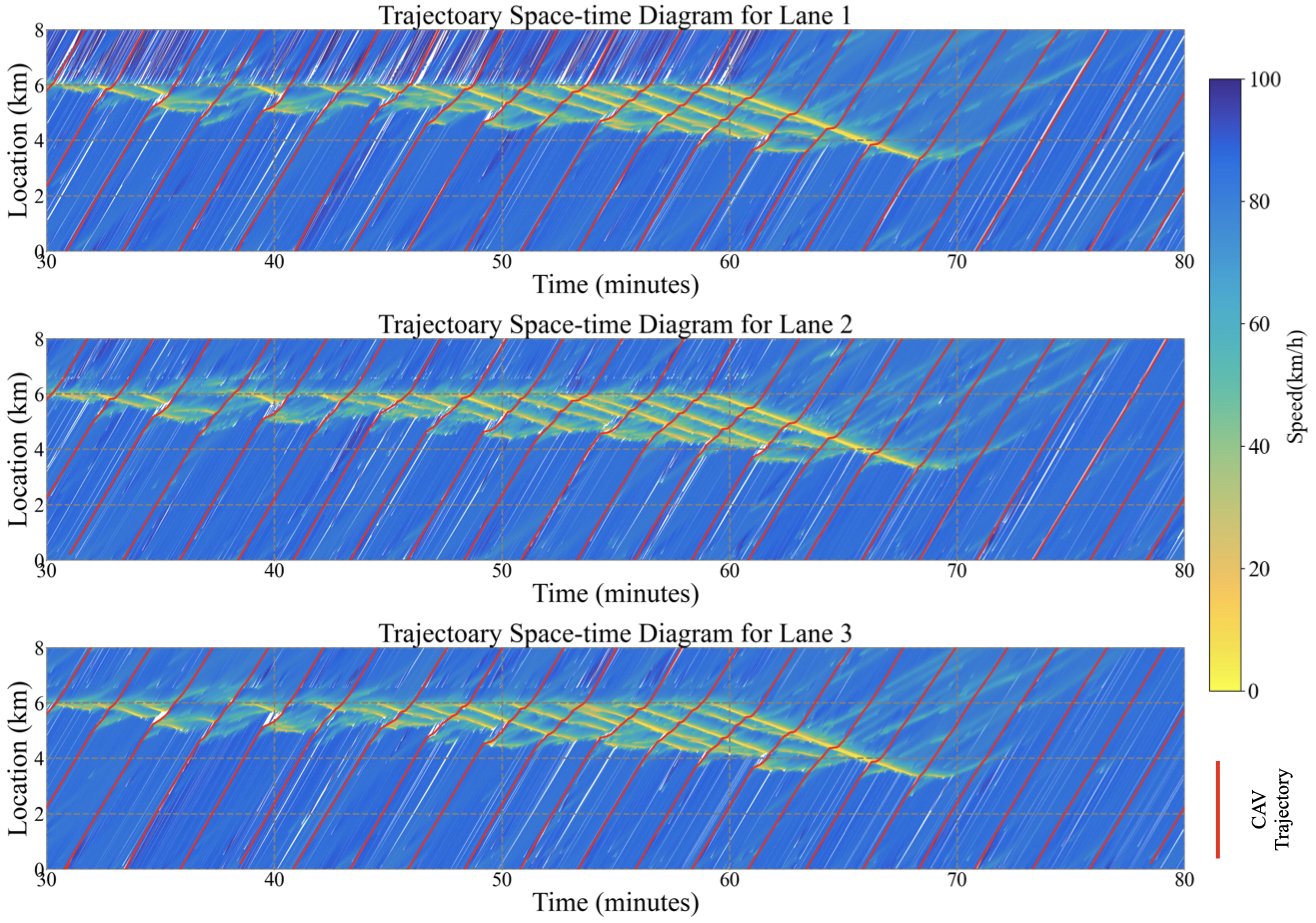}
      \caption{Vehicle trajectory space-time diagram for the controlled scenario using GN-LQR controller with $N=3$ (the color of each trajectory point reflects the speed of the vehicle at the time, and the CAV-platoons trajectories are labeled with red lines): [top] the left lane, [middle] the middle lane, [bottom] the right lane.}
  \label{f:micro_lqr_traj}
  \end{figure}

Compared to the uncontrolled case, the implementation of the GN-LQR controller results in an increase in the MS of traffic to 79.43 by reducing the TTT from 1,089 to 996 while the TTD only changes by a small amount. Figure \ref{f:micro_lqr_traj} presents the space-time diagram for the trajectories in the simulation for the scenario with GN-LQR, which helps develop a qualitative understanding of the performance of the controller in a realistic setting. 
Notice that in comparison to the uncontrolled case, in this case, the jam only extends to around the $4$ km mark. The slowdown of CAV-platoons can be observed from the reduction in the slope of the red lines on the plot (depicting the trajectories of the CAV-platoons) which occurs due to the control as well as naturally when the platoons meet the jam wave created by the bottleneck. While the controlled slowdown of CAV-platoons still results in a slowdown of vehicles upstream of the platoons, the resulting jam waves are much less severe and shorter resulting in less disruption to upstream traffic which recovers quickly. This causes the overall jam to reduce in size and increases the speed of traffic in general. This is similar to the observations made in the case of the macroscopic simulation-based analysis and provides preliminary confirmation of the usability of the controller in a realistic setting.

\begin{figure}
      \centering
      \includegraphics[width=1\textwidth]{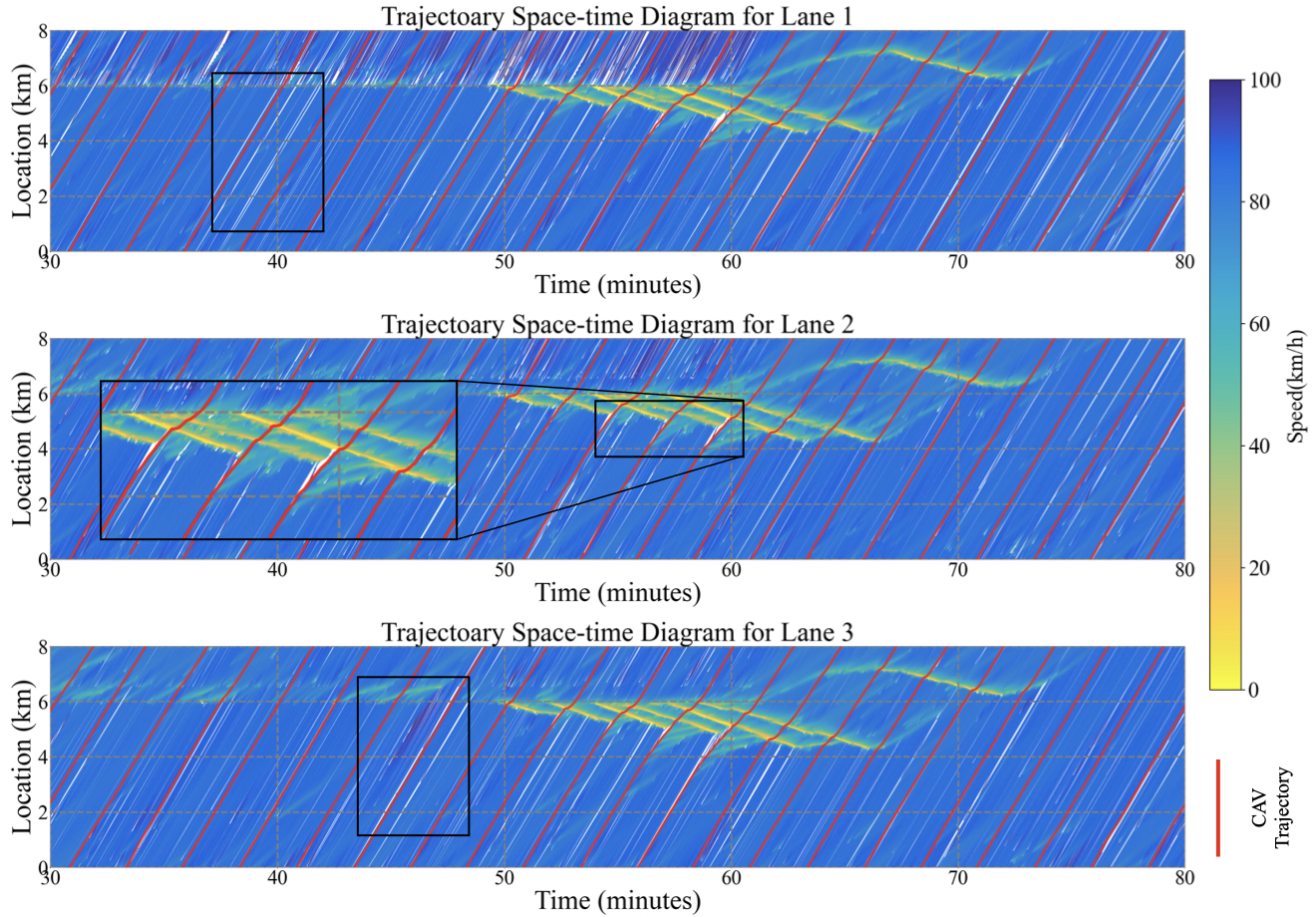}
      \caption{Vehicle trajectory space-time diagram for the controlled scenario using GN-LQRP controller (the color of each trajectory point reflects the speed of the vehicle at the time, and the CAV-platoons trajectories are labeled with red lines): [top] the left lane, [middle] the middle lane, [bottom] the right lane. }
      \label{f:micro_lqrp_traj}
  \end{figure}

For the implementation of GN-LQRP, we consider three different values of $N$ namely $50, 30$, and $10$ with all other parameter values the same in Section \ref{s:comparison}. The evaluation metrics for the three cases are presented in Table \ref{t:micro_comp}. While $N=50$ is determined as the best tuning value for the controller in the macroscopic setting, it is found that it is not the best value in the microscopic setting where it leads to stopping of vehicles more upstream of the bottleneck sometimes causing significant jam waves that lead up to the upstream end of the highway stretch (see Figure \ref{f:micro_lqrp50_traj} in Appendix \ref{s:trajectories}). While GN-LQRP with $N=50$ also results in a slowdown of vehicles more upstream of the bottleneck in the macroscopic setting, it does not lead to severe jams in that case. This difference between the macroscopic and microscopic scenarios for the GN-LQRP controller can be explained by the inherent differences in the impact of the vehicle slowdowns between the macroscopic and microscopic models. The TTD in this case is also smaller because of a reduction in the number of vehicles entering the stretch due to jams at the upstream end of the highway stretch. As found in Section \ref{s:LQR_N}, reducing $N$ results in a slowdown of vehicles to less upstream of the bottleneck. Therefore, here we also test the controller with reduced values of $N=10$ and $N=30$. It is observed that smaller $N$ does prevent jams at the upstream end of the stretch restoring the value of TTD as compared to the case with $N=50$. In fact, GN-LQRP with $N=10$ is also able to outperform GN-LQR in the microscopic setting. From Figure \ref{f:micro_lqrp_traj}, it can be observed that as compared to GN-LQR, GN-LQRP with $N=10$ noticeably results in more gradual changes in the speed of traffic close to the bottleneck which results in reduced disruptions in upstream traffic from the controlled slowdowns. This is expected from GN-LQRP as it inherently penalizes abrupt changes in control speeds which are less realistic and provides optimal control under restricted speed changes.

 Finally, the PI- and MPC-based controllers also result in an improvement in the state of traffic over the uncontrolled case by reducing the TTT and as a result the MS as seen from Table \ref{t:micro_comp}. The plots of vehicle trajectories for the PI- and MPC-based controllers are presented in Appendix \ref{s:trajectories}. Between the two controllers, the MPC-based controller performs better and almost comparably with GN-LQRP while the PI-based controller performs worse than GN-LQR in microsimulation. Both the controllers show limited changes in CAV-platoon speeds (prior to joining the jam waves) as seen from the trajectory plots which is due to the lower bound on control speeds. MPC is able to use the limited changes to reduce the size of the jam more significantly as also observed from Figure \ref{f:ms_density}.

\section{Conclusions and future direction} \label{s:conclusion}
From the previous analysis, we have some preliminary suggestions regarding the questions posed in Section \ref{s:study_objectives} which are as follows:
\begin{enumerate}
    \item \textit{A1:} Both GN-LQR and GN-LQRP controllers are able to reduce the negative effects of fixed bottlenecks on the highway stretch in both macroscopic and microscopic traffic settings. The performance of the GN-LQR controller is comparable to the MPC-based controller (with and without a lower bound on control speeds) and the PI-based controller (with a lower bound on control speeds) in macrosimulation. The PI-based controller without a lower bound on the control speed does not improve the condition of traffic.

    \item \textit{A2:} Both LQR-based controllers outperform the MPC-based controller in terms of average computation time for each controller run. Based on obtained results, LQR-based controllers are at least $30$ times faster than the MPC-based controller (comparing computation time between GN-LQRP and MPC with a lower bound). The PI-based controller is the fastest as it only requires the offline computation of the gains. However, this is contingent upon the existence of a reliable way to compute the gains offline.

    \item \textit{A3:} The controls obtained from all the controllers are plausible in terms of maximum acceleration/deceleration requirements for vehicles to achieve the prescribed speeds. However, accounting for human reaction times, the PI-based controller (with a lower bound on the control speed) and the GN-LQRP controller offer the safest and most realistically achievable controls as they tend to gradually vary the speeds of the CAV-platoons unlike the MPC-based controller and the GN-LQR controller which can jump from the upper to the lower bound of control speeds within one time step.
    
    \item \textit{A4:} We obtain interesting insights into the impact of various parameters on the performance of the controllers in a macroscopic simulation setting. The GN-LQR controller performs well at lower values of horizon lengths up to $N=20$ time steps. Beyond $N=20$, we observe congestion and spillback caused at the upstream end of the highway which is not desirable from a traffic control perspective. GN-LQRP on the other hand performs worse than GN-LQR for smaller values of $N$ and achieves its peak performance around $N=50$ time steps beyond which the performance degrades. The ideal number of iterations for the algorithms is $1$ and a larger number of iterations adversely affects the performance. The magnitude of the state error weight matrix $\m Q$ needs to be much larger than the weight matrix on input error term $\m R$. A larger magnitude of the weight matrix $\m R'$ for the penalty term in the GN-LQRP controller results in worse performance in terms of $\textrm{MS}$ at the same values of $N$ due to restricted speed changes. However, larger values of $\m R'$ result in less varying/safer control.

    \item \textit{A5:} Both GN-LQR and GN-LQRP controllers improve the MS metric of traffic over the uncontrolled case in the microscopic traffic simulation by reducing the speed of the traffic approaching the bottleneck resulting in a reduction in the length of the formed jam waves. While the same tuning parameters as the macrosimulation analysis for both the proposed controllers improve the traffic state in microsimulation as well, it is observed that GN-LQRP works better with smaller values of $N=10$ in microsimulation. Larger values of $N$ result in the slowing down of vehicles a long way upstream of the bottleneck resulting in comparatively worse performance than when the CAVs only slow down close to the bottleneck which takes place with a smaller $N$. Differences in the optimal tuning of controllers between the macroscopic and microscopic simulations are expected due to the inherent differences between the models and the difficulty associated with the precise determination of macroscopic model parameters used within the controller which can affect the traffic dynamics. The existing controllers are also observed to work well in microsimulation. In the current study, GN-LQRP ($N=10$) and MPC (with a lower bound on the control speed) marginally outperform other controllers.
\end{enumerate}
 The current analysis compared the different controllers at the same random seed value for the microsimulation. While understanding the impact of simulator stochasticity on the controllers' performance is out of the scope of the current work which mainly focuses on proposing and analyzing new controllers for traffic control and performing a preliminary test on their usability in microsimulation, performing multiple runs of the simulator with different seed values to account for stochasticity and analyzing them is necessary to develop a deeper understanding of the effectiveness of CAV-based control in the real world. There is also scope for a more in-depth analysis of CAV-based control under different scenarios such as in the presence of different jam/bottleneck triggering factors including but not limited to the existence of curvature, slope, lane-drop, tunnels, bridges, imposed speed limits, or ramps. There are several potential directions of improvement for the controllers mainly with regard to application in the microscopic simulation case which would further improve their performance in the real world. These include the extension of the control approach to account for uncertainty in parameter estimation for the state-space equation used in the controller as well as accounting for the lack of compliance of the CAVs to the control inputs. These changes would be a step in the direction of robust traffic control using CAVs. Also, additional flexibility in the form of lane-wise control of CAVs can be considered wherein the CAVs are no longer restricted to travel side-by-side but can block the traffic in individual lanes in coordination with other CAVs to achieve optimal control. This would require a lane-wise macroscopic traffic model similar to one in \cite{bekiaris2017highway} to define the state-space model for the controller. Besides, the current work can also be extended to large-scale road networks and consideration of other forms of control such as ramp metering and variable speed limits which can be incorporated into the framework as inputs to the system thus allowing for integrated control.

\bibliographystyle{IEEEtran}
\bibliography{biblio}

\appendices

\section{LQR optimization problem with a penalty on control input changes}
\label{a:GN-LQRP}
 This section describes the formulation of the modified LQR optimization problem which penalizes changes in control inputs over consecutive time steps. To explain the idea of this controller, we use the equations for a standard linear system of the form
\begin{align}
    \m x[k+1]&=\m A\m x[k] + \m B \m u[k], \nonumber\\
    \m u[k] &= \m u[k-1] + \Delta \m u[k], \label{e:control_change}
\end{align}
where \eqref{e:control_change} simply describes the current input in terms of the previous input and the change in input. This system can also be written as follows:
\begin{align}\label{e:aug_primary}
    \begin{bmatrix}\m x[k+1] \\ \m u[k]\end{bmatrix} = \begin{bmatrix}\m A & \m B\\ \m 0 & \m I \end{bmatrix}\begin{bmatrix}\m x[k] \\ \m u[k-1]\end{bmatrix} + \begin{bmatrix}\m B \\ \m I\end{bmatrix}\Delta \m u[k].
\end{align}
By defining new augmented vectors and matrices,
\begin{align}\label{e:augmented_matrices}
    & \m x'[k]:=\begin{bmatrix}
        \m x[k]\\ \m u[k-1]
    \end{bmatrix}, \quad
    \m A' := \begin{bmatrix}\m A & \m B\\ \m 0 & \m I \end{bmatrix},\quad
    \m B' := \begin{bmatrix}\m B \\ \m I\end{bmatrix},\\ & \m u'[k] := \Delta \m u[k],
\end{align}
we can write \eqref{e:aug_primary} as follows:
\begin{align}\label{eq:augmented_state_space}
    \m x'[k+1] &= \m A' \m x'[k] + \m B' \m u'[k].
\end{align}

Since \eqref{eq:augmented_state_space} resembles a standard linear system, we can write a new optimization problem in the LQR framework which regulates both the states and the change in control inputs (instead of the control inputs directly) around the zero point. The objective function of this problem is defined as follows:
\begin{align}
J'(\m x'[k],\m u'[k]) &:= \m x'[k]^T \m Q' \m x'[k] + \m u'[k]^T \m R' \m u'[k],
\label{e:new_LQR_objective}
\end{align}
where
\begin{align}
\m Q' = \begin{bmatrix}
\m Q & \m 0\\
\m 0 & \m R
\end{bmatrix},
\label{e:augQ}
\end{align}
and $\m R'\in\mathbb{R}^{(n_u\times n_u)}$ is the weight matrix for the penalty on control input changes.

The same idea can be applied to regulate the states of the nonlinear system \eqref{SSM} around a predefined equilibrium point while keeping the change in the control inputs to a minimum. The GN-LQR algorithm (Algorithm \ref{alg:cap}) can be applied for this purpose with some modifications. The modified algorithm is presented as Algorithm \ref{alg:cap2}. 

For Algorithm \ref{alg:cap2}, the augmented states and control inputs and the corresponding equilibrium points are  defined in the same way as in \eqref{e:augmented_matrices} which are further used to obtain the stacked augmented matrices corresponding to $\m X, \m U, \m X^{\m \ast}, \m U^{\m \ast}$. The augmented state-space matrices are defined using the linearized state-space matrices \eqref{e:linearized_state_space} as follows:
\begin{align}\label{e:augA}
    \hat{\m A}' := \begin{bmatrix}\hat{\m A} & \hat{\m B}\\ \m 0 & \m I \end{bmatrix},\quad
    \hat{\m B}' := \begin{bmatrix}\hat{\m B} \\ \m I\end{bmatrix}.
\end{align}

\section{PI-based controller implementation}
\label{a:PI}
The PI-based controller implemented in this work is based on that presented in \cite{piacentini2019multiple}. The control law is given as follows:
\begin{equation}
\label{e:PI_speed_update}
    u_j[k]=\bar{v}_j[k-1]+K_P(e_j[k]-e_j[k-1])+K_I e_j[k],
\end{equation}
where $K_P$ and $K_I$ are the controller gains, and $e[k]$ is the controller error which defined as follows:
\begin{equation}
    e_j[k]=\hat{\rho}-\bar{\rho}_j[k],
\end{equation}
where $\bar{\rho}_j[k]$ is the average density over segments downstream of CAV-platoon $j$ and upstream of the fixed bottleneck on the highway stretch, and $\hat{\rho}$ is called the density set-point which is supposed to be the ideal value of $\bar{\rho}_j[k]$ and is set equal to the critical density $\rho_c$. Here the average density is calculated using only the segments whose density is above a certain threshold value which in this case is set to the critical density. So if none of the segments between the CAV-platoon and the fixed bottleneck have density above $\rho_c$, then the $e_j[k]$ is undefined (since there are no segments to calculate it over) and $\bar{v}_j[k]=\bar{v}_j[k-1]$.

 In this work, the optimal gains for the PI-based controller are obtained by setting up a nonlinear optimization problem with the objective of maximizing the MS. The \texttt{fmincon()} solver of MATLAB which implements the interior point algorithm is used to solve the nonlinear optimization problem as a minimization problem. The state evolution steps for the full duration of the simulation are set as constraints within \texttt{fmincon()} while the objective is set to the negative of MS. The bounds for the gain values are set to $[-10,10]$ for both gains which are found to be sufficient. The solver is initialized with the solution values $0.8$ and $1.6$ which are obtained from \cite{piacentini2019multiple}.

\section{MPC-based controller implementation}
\label{a:MPC}
The MPC-based controller is implemented based on the implementation in \cite{piacentini2019highway}. The corresponding optimization problem is given as follows:
\begin{align}
    \hspace{-4mm}\min_{\m \rho[k],\m u[k]}\quad &\beta_1T\sum_{h=k}^{k+N_P}\sum_{i=1}^{N_L}{L_i\rho_i[h]}-\beta_2\sum_{h=k}^{k+N_P}\phi_{\bar{i}}[h] -\beta_3\sum_{h=k}^{k+N_P}|\rho_{\bar{i}}[h]-\rho_c|\nonumber\\
    \hspace{-4mm}s.t. \quad & \m u^{min}\le \m u[h] \le \m u^{max} \hspace{2mm}\textrm{for}\hspace{2mm} h=k,\dots,k+N_P,
\end{align}
where $k$ is the current time step, $T$ is the duration of a time step, $N_P$ is the prediction horizon which is set to 20-time steps as in \cite{piacentini2019highway}, $N$ is the number of segments on the considered highway stretch, $L_i$ is the length of segment $i$, $\rho_i[h]$ is the density of segment $i$ at time step $h$, $\rho_{\bar{i}}[h]$ is the density of the bottleneck segment at time step $h$, $\phi_{\bar{i}}[h]$ is the outflow from the bottleneck segment at time step $h$, $\beta_1, \beta_2,$ and $\beta_3$ are the objective weights set to $0.1, 0.1$ and $0.8$ respectively as in \cite{piacentini2019highway}, and $\m u^{min}$ and $\m u^{max}$ are the upper and lower bound on the control speeds. While we only control the CAV-platoon speeds explicitly in this optimization, the density of highway segments is also a variable since it changes with the value of the control speed.
The above optimization problem is solved using the interior point algorithm implemented through the $\texttt{fmincon()}$ solver of MATLAB. 

\section{Car-following model}
\label{a:car_following}
The car-following model used in the microsimulation is the intelligent driver car-following model \cite{kesting2008calibrating}, shown in Equations \eqref{e:idm1}-\eqref{e:idm3}, the detailed parameters are listed in Table \ref{t:idm}. In TransModeler, only the desired time gap $T$ and the free acceleration exponent $\delta$ are editable.
\begin{align}
& \dot{x}_\alpha=\frac{\mathrm{d} x_\alpha}{\mathrm{d} t}=v_\alpha, \label{e:idm1} \\
& \dot{v}_\alpha=\frac{\mathrm{d} v_\alpha}{\mathrm{d} t}=a\left(1-\left(\frac{v_\alpha}{v_0}\right)^\delta-\left(\frac{s^*\left(v_\alpha, \Delta v_\alpha\right)}{s_\alpha}\right)^2\right),\label{e:idm2} \\
&s^*\left(v_\alpha, \Delta v_\alpha\right)=s_0+v_\alpha T+\frac{v_\alpha \Delta v_\alpha}{2 \sqrt{a b}}.
\label{e:idm3}
\end{align}

\begin{table}[h]
\centering
\caption{Parameter setting for the Intelligent Driver Car-following Model in TransModeler.}
\label{t:idm}
\begin{tabular}{|c|c|c|l|}
\hline
Parameters & Value & Unit & Physical Meaning \\ \hline
$a$  & 2.0 & $m/s^2$&  the maximum vehicle acceleration            \\\hline
$b$  & 1.0 & $m/s^2$ &  comfortable braking deceleration              \\\hline
$s_0$  & 0.0 & $m$&  minimum (safety) stopping distance             \\\hline
$T$  & 1.2 &$s$ &  desired time gap               \\\hline
$\delta$  & 6.0 &- & free acceleration exponent                \\\hline
$v_0$  & 100& $km/h$ &  desired (or free-flow) speed              \\\hline
\end{tabular}
\end{table}

\vspace{4cm}
\section{Trajectory space-time diagrams for microsimulation}
\label{s:trajectories}

  \begin{figure}[H]
      \centering
    \includegraphics[width=1\textwidth]{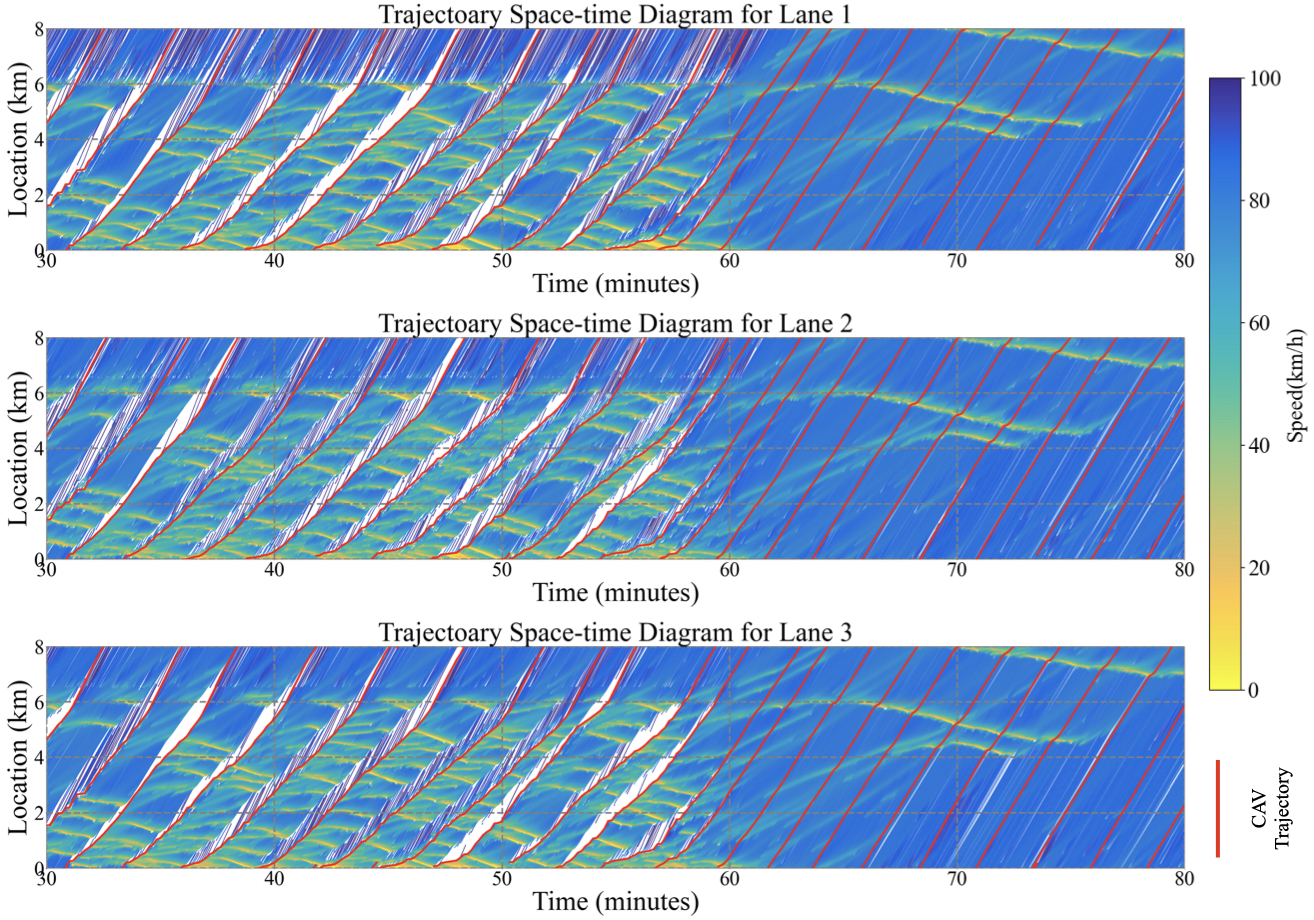}
      \caption{Vehicle trajectory space-time diagram for GN-LQRP ($N=50$) (the color of each trajectory point reflects the speed of the vehicle at the time): [top] the left lane, [middle] the middle lane, [bottom] the right lane.}
      \label{f:micro_lqrp50_traj}
  \end{figure}
  
  \begin{figure}
      \centering
      \includegraphics[width=1\textwidth]{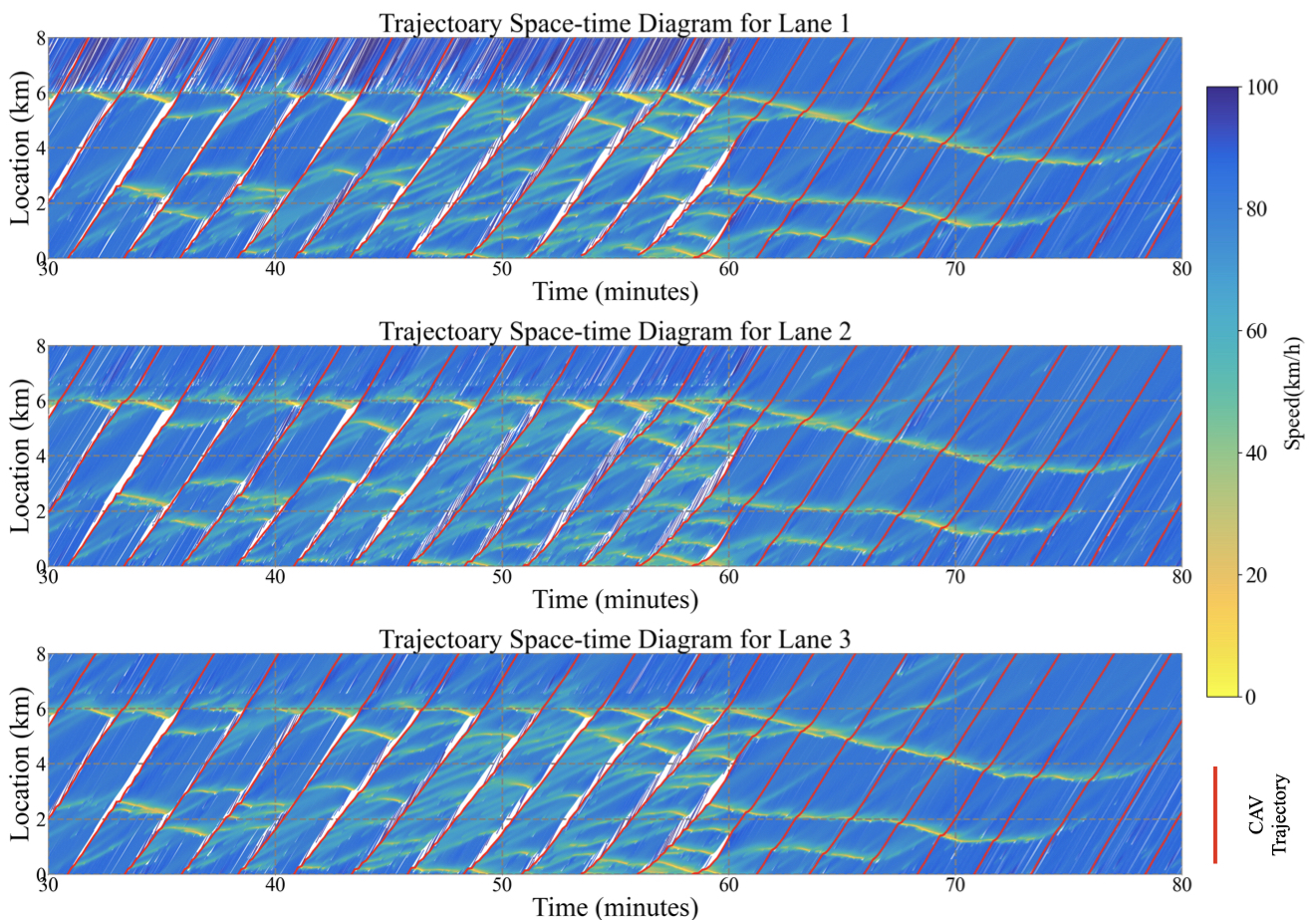}
      \caption{Vehicle trajectory space-time diagram for GN-LQRP ($N=30$) (the color of each trajectory point reflects the speed of the vehicle at the time): [top] the left lane, [middle] the middle lane, [bottom] the right lane.}
  \end{figure}
  \begin{figure}
      \centering
      \includegraphics[width=1\textwidth]{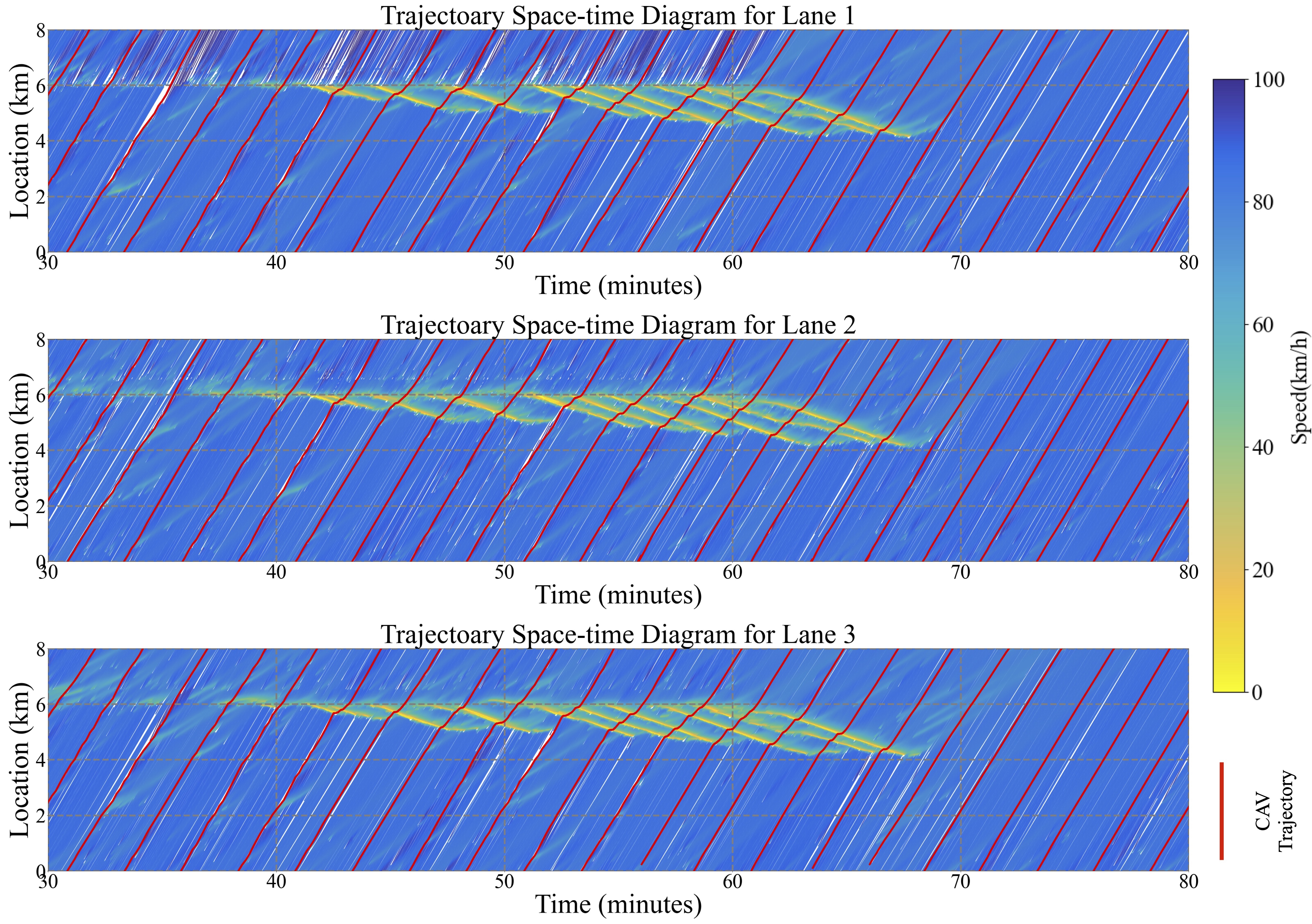}
      \caption{Vehicle trajectory space-time diagram for MPC (the color of each trajectory point reflects the speed of the vehicle at the time): [top] the left lane, [middle] the middle lane, [bottom] the right lane.}
  \end{figure}
  \begin{figure}
      \centering
      \includegraphics[width=1\textwidth]{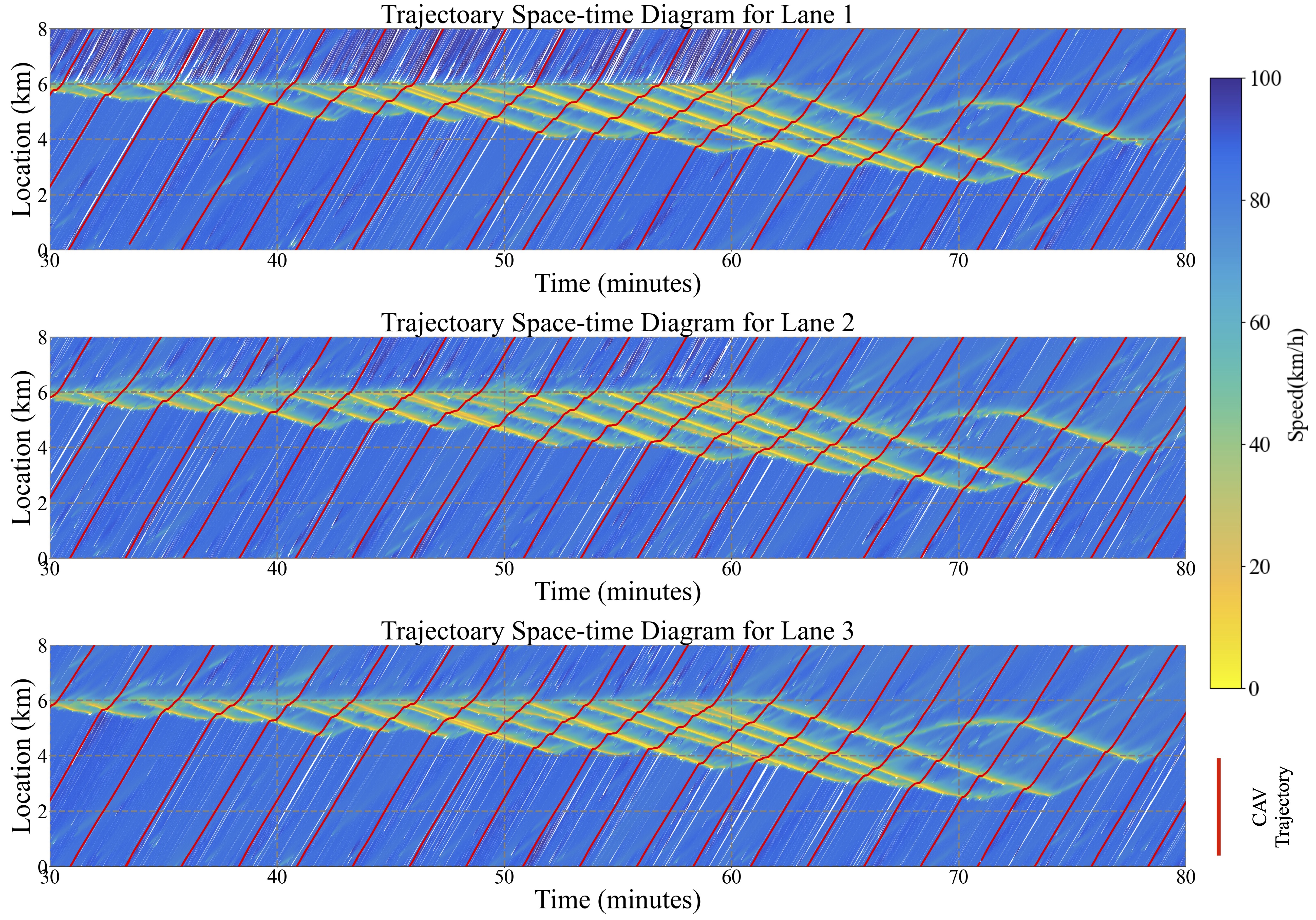}
      \caption{Vehicle trajectory space-time diagram for PI (the color of each trajectory point reflects the speed of the vehicle at the time): [top] the left lane, [middle] the middle lane, [bottom] the right lane.}
  \end{figure}
\end{document}